 \documentclass{article}

\usepackage{epsfig} 
\usepackage{amsmath}
\usepackage{amsfonts}
\usepackage{graphicx}
\usepackage{cite}

 \usepackage{amssymb}

\date{\today}

\newcommand{\insertplot}[5]{\begin{figure}
 \hfill\hbox to 0.05in{\vbox to #5in{\vfill
 \inputplot{#1}{#4}{#5}}\hfill}
 \hfill\vspace{-.1in}
 \caption{#2}\label{#3}
 \end{figure}}
 \newcommand{\inputplot}[3]{
 \special{ps: plotfile #1}
}
\newcounter{fig}   

\numberwithin{equation}{section}

\newcommand{\ee}{\end{equation}}
\newcommand{\eea}{\end{eqnarray}}
\newcommand{\be}{\begin{equation}}
\newcommand{\bea}{\begin{eqnarray}}

\begin{document}


\title{\bf \Large Kerr black holes with Proca hair}

 \author{
{\large Carlos Herdeiro}\footnote{herdeiro@ua.pt}, \
{\large Eugen Radu}\footnote{eugen.radu@ua.pt} \
and
{\large Helgi R\'unarsson}\footnote{helgi.runarsson@gmail.com}
\\ 
\\
{\small Departamento de F\'\i sica da Universidade de Aveiro and} \\ 
{\small Center for Research and Development in Mathematics and Applications (CIDMA)} \\ 
{\small   Campus de Santiago, 3810-183 Aveiro, Portugal}
}


\date{March 2016}
\maketitle

\begin{abstract}
Bekenstein proved that in Einstein's gravity minimally coupled to one (or many) real, Abelian, Proca field, stationary black holes (BHs) cannot have Proca hair. Dropping Bekenstein's assumption that matter inherits spacetime symmetries, we show this model admits asymptotically flat, stationary, axi-symmetric, regular on and outside an event horizon BHs with Proca hair, for an even number of real  (or an arbitrary number of complex) Proca fields. To establish it, we start by showing that a test, complex Proca field can form bound states, with real frequency, around Kerr BHs: \textit{stationary Proca  clouds}. These states exist at the threshold of superradiance.  It was conjectured in~\cite{Herdeiro:2014goa,Herdeiro:2014ima}, that the existence of such clouds at the linear level implies the existence of a new family of BH solutions at the non-linear level. We confirm this expectation and explicitly construct examples of such \textit{Kerr black holes with Proca hair} (KBHsPH). For a single complex Proca field, these  BHs form a countable number of families with three continuous parameters (ADM mass, ADM angular momentum and Noether charge). They branch off from the Kerr solutions that can support stationary Proca clouds and reduce to Proca stars~\cite{Brito:2015pxa} when the horizon size vanishes. We present the domain of existence of one family of KBHsPH, as well as its phase space in terms of ADM quantities. Some physical properties of the solutions are discussed; in particular, and in contrast with Kerr BHs with scalar hair, some spacetime regions can be counter-rotating with respect to the horizon. 
We further establish a no-Proca-hair theorem for static, spherically symmetric BHs but allowing the complex Proca field to have a harmonic time dependence, which shows BHs with Proca hair in this model require rotation and have no static limit. KBHsPH are also disconnected from Kerr-Newman BHs with a real, massless vector field.

\end{abstract}





 
 \newpage

\section{Introduction}
 
In vacuum General Relativity (GR) black holes (BH) are remarkably simple. The Carter-Robinson theorem~\cite{Carter:1971zc,Robinson:1975bv}, supplemented by the rigidity theorem~\cite{Hawking:1971vc,Racz:1995nh}, established that asymptotically flat, stationary, non-singular (on and outside an event horizon)  \textit{vacuum} BHs of GR have \textit{only} 
two degrees of freedom -- see~\cite{Chrusciel:2012jk} for a review.  

The most general BH solution in this context is the Kerr metric~\cite{Kerr:1963ud} and the two degrees of freedom are the ADM mass, $M$, and angular momentum, $J$, both of which can be determined by an observer at infinity. 

The natural question of how this result generalizes in the presence of matter led to the \textit{no-hair} hypothesis~\cite{Ruffini:1971bza}: regardless of the matter involved, the end-point of gravitational collapse -- in GR and in an astrophysical context -- is characterized solely by conserved charges associated to Gauss laws, including $M,J$, and no further parameters (\textit{hair}). Thus, an observer at infinity should be able to fully compute all relevant ``charges" of an equilibrium BH.

Evidence in favour of this hypothesis has been presented in terms of \textit{no-hair theorems} for particular matter models in GR. A collection of such theorems for the much studied case of scalar matter can be found in the recent review~\cite{Herdeiro:2015waa}. Of relevance for the present paper, Bekenstein established a no-Proca hair theorem for stationary BH solutions of Einstein's gravity minimally coupled to one (or more) real, Abelian Proca field~\cite{Bekenstein:1971hc,Bekenstein:1972ky}, which will be reviewed in Section~\ref{sec_nohair1}.

Evidence against the no-hair hypothesis in asymptotically flat spacetimes, on the other hand, has been presented in the form of hairy BH solutions, starting with the pioneering examples in Yang-Mills theory~\cite{Volkov:1998cc} (see also the reviews~\cite{Bizon:1994dh,Bekenstein:1996pn,Herdeiro:2015waa,Volkov:2016ehx}). 
Such counter-examples, however, typically either 
1) violate some energy condition ($e.g.$~\cite{Nucamendi:1995ex,Bechmann:1995sa,Anabalon:2013qua,Anabalon:2012ih,Cadoni:2015gfa}); or 
2) have non-minimal couplings between matter and geometry ($e.g.$~\cite{Bekenstein:1975ts,Bekenstein:1974sf,BBM,Sotiriou:2014pfa,Sotiriou:2013qea,Babichev:2013cya}); or 
3) have non-canonical/non-linear kinetic terms ($e.g.$~\cite{Luckock:1986tr,Luckock,Bronnikov:2005gm,Radu:2011uj}); or 
4) the hair is not independent of other fields, such as an electromagnetic field (secondary hair, $e.g.$~\cite{Gibbons:1987ps,Gibbons:1982ih}); or 
5) involve higher curvature terms 
($e.g$~\cite{Kanti:1995vq,Ayzenberg:2014aka,Kleihaus:2011tg,Kleihaus:2015aje,Pani:2011gy,Pani:2009wy,Alexander:2009tp,Yunes:2009hc}); or 
6) several of the above. It is unclear, moreover, if any of these counter-examples violates the \textit{dynamical spirit} of the no-hair hypothesis; that is, if there are dynamically stable hairy BHs that can be the end-point (or be sufficiently long lived) in a dynamical evolution.

\bigskip

In a qualitatively novel development, a class of BH solutions with scalar hair was found in 2014 bifurcating from the Kerr metric~\cite{Herdeiro:2014goa}: \textit{Kerr BHs with scalar hair} (KBHsSH).
 These are solutions of the simple model
 \begin{equation}
\label{actionscalar}
S=\int  d^4x \sqrt{-g}\left[ \frac{R}{16\pi G}
   -\frac{g^{\alpha\beta}}{2} \left( \Psi_{, \, \alpha}^* \Psi_{, \, \beta} + \Psi _
{, \, \beta}^* \Psi _{, \, \alpha} \right) - \mu^2 \Psi^*\Psi
 \right]  \ ,
\end{equation}
that 1) obey all energy conditions; 2) have minimal couplings with the geometry; 3) have canonical kinetic terms; 4) have an independent (primary) hair; 5) exist in GR, without higher curvature terms. KBHsSH, moreover, are asymptotically flat, regular on and outside the event horizon, reduce to (specific) Kerr solutions in the limit of vanishing hair, and to gravitating solitons known as \textit{boson stars}~\cite{Schunck:2003kk,Liebling:2012fv} in the limit of vanishing horizon. The scalar hair is described by an independent  conserved Noether charge \textit{but without} an associated Gauss law. Thus, an observer at infinity cannot determine this charge -- which must be computed by a volume integral -- and hence does not have access to all the relevant spacetime charges.  

The matter content for the original example in~\cite{Herdeiro:2014goa} (see also~\cite{Herdeiro:2015gia}) was a massive complex scalar field, $cf.$~\eqref{actionscalar}. In GR minimally coupled to this type of matter the Kerr BH is a solution, together with a vanishing scalar field, but it is \textit{unstable} against superradiance~\cite{Press:1972zz,Damour:1976kh,Brito:2015oca}. At the threshold of the instability, there are bound states of the scalar field on the Kerr background, found in a test field analysis, corresponding to linear hair. The existence of these \textit{stationary scalar clouds}~\cite{Hod:2012px,Hod:2013zza,Herdeiro:2014goa,Hod:2014baa,Benone:2014ssa,Hod:2014npa,Hod:2015ota} determines the bifurcation point of the hairy solutions from vacuum Kerr. Moreover, since the latter solution is unstable against superradiant scalar perturbations, there is an expectation that the BHs of~\cite{Herdeiro:2014goa} play a role in the non-linear development of the instability and can effectively form dynamically, thus providing a true counter-example to the physical implications of the no-hair hypothesis -- see~\cite{Sanchis-Gual:2015lje,Bosch:2016vcp} for recent discussions of the non-linear development of superradiant instabilities into hairy BHs. 

\bigskip

The connection between KBHsSH and superradiance led to the suggestion that, underlying the example of KBHsSH, there is a more general mechanism~\cite{Herdeiro:2014goa,Herdeiro:2014ima} (see also~\cite{Herdeiro:2015waa,Herdeiro:2015gia}): 
\newpage
{\bf Conjecture:}
\begin{description}
\item[1)] If a ``hairless" stationary BH spacetime $(\mathcal{M}_0,{\bf g_0})$ is afflicted by superradiant instabilities triggered by a given test field $\mathcal{F}$;
\item[2)] If the field modes at the threshold of the instability (zero modes), $\mathcal{F}_t$, yield an energy-momentum tensor $\mathcal{T}(\mathcal{F}_t,\mathcal{F}_t)$ which is time-independent $\mathcal{L}_{\bf k}\mathcal{T}(\mathcal{F}_t,\mathcal{F}_t)=0$, where ${\bf k}$ is the time-like Killing vector field (at infinity) that preserves the metric $\mathcal{L}_{\bf k}{\bf g_0}=0$;
\item[Then:] there is a new family of stationary BH ``hairy solutions" bifurcating from $(\mathcal{M}_0,{\bf g_0})$, denoted by $(\mathcal{M}_\mathcal{F},{\bf g_\mathcal{F}})$. Actually, $(\mathcal{M}_\mathcal{F},{\bf g_\mathcal{F}})$ may be a countable set of families.
\end{description}
In the case of KBHsSH, one encounters a family with three continuous and two discrete degrees of freedom. The former are the ADM mass and angular momentum $(M,J)$ and the Noether charge $Q$; the latter, which define a countable set of families, are the node number, $n\in \mathbb{N}_0$ and the azimuthal harmonic index, $m\in \mathbb{Z}$ of the scalar field. A formal proof of the existence of these solutions was recently reported in~\cite{Chodosh:2015oma}. KBHsSH were generalized to include self-interactions of the scalar field in~\cite{Herdeiro:2015tia} and to scalar-tensor gravity in~\cite{Kleihaus:2015iea}.

\bigskip

As further evidence for the above conjecture we consider, in this paper, 
Einstein's gravity minimally coupled to Abelian Proca fields, hereafter referred simply as Proca fields\footnote{Gravitating \textit{non-Abelian} ($SU(2)$) Proca fields have been studied in \cite{Greene:1992fw}, wherein spherically symmetric solitons and BHs have been discussed. The properties of these solutions are rather distinct from the solutions discussed in this paper and, moreover, the former have not been generalized to include rotation.}. 
Massive Proca fields trigger, in much the same way as massive scalar fields, 
superradiant instabilities of Kerr BHs -- see~\cite{Pani:2012vp,Pani:2012bp,Witek:2012tr} 
for recent studies of Proca-induced superradiant instabilities in asymptotically flat BHs. 
Firstly, we shall perform a test field analysis of a Proca field on the Kerr background. We observe that, at the threshold of the unstable modes, one can find \textit{stationary Proca clouds}. 
If the Proca field is complex, moreover, the energy-momentum tensor sourced by these stationary clouds is time-independent. Hence, we are in the conditions of the above conjecture. Secondly, we address the fully non-linear system of a complex-Proca field minimally coupled to GR and construct stationary BH solutions which are the non-linear realization of the aforementioned stationary Proca clouds: \textit{Kerr BHs with Proca hair} (KBHsPH). When the horizon of these BHs vanishes, the solutions reduce to the rotating Proca stars recently constructed in~\cite{Brito:2015pxa}. These are vector boson stars which share many of the properties of the scalar boson stars that have been studied for decades~\cite{Schunck:2003kk,Liebling:2012fv}.

\bigskip

The introduction of the mass term for the vector fields is central for the existence of KBHsPH, since it is crucial for both the existence of the stationary Proca clouds and Proca stars. In the (Proca field) massless limit these BHs trivialize; they are not connected to Kerr-Newman BHs. The presence of such mass terms implies that there is no Gauss law associated to the vector field;  massive fields have no Gauss law since there is no flux conservation. Indeed, in asymptotically flat spacetimes, a massive field which decays towards spatial infinity will do so exponentially. Thus, the integral of its flux density over a sphere at infinity will necessarily vanish. 

This does not mean, however, that massive fields cannot be \textit{locally} conserved. Both the complex Proca field and the complex, massive scalar field enjoy a $U(1)$ global symmetry which implies a conserved current and a conserved Noether charge. There is a local continuity equation, but no Gauss law. Thus, according to the no-hair hypothesis there should be no Proca hair around stationary BHs. Here, however, we show that there can be. And again, an observer at infinity does not have access to all relevant spacetime charges.

\bigskip

This paper is organized as follows. In Section~\ref{sec_model} we exhibit the Einstein--complex-Proca model and its basic properties. In Section~\ref{sec_nohair} we review the classic no-Proca-hair theorem by Bekenstein~\cite{Bekenstein:1971hc,Bekenstein:1972ky} and also present a novel no-Proca-hair theorem applying to \textit{spherically symmetric} solutions and allowing the Proca field to have a harmonic time dependence. The latter is a generalization of the theorem presented in~\cite{Pena:1997cy} for the scalar case and it establishes that rotation is crucial for the existence of KBHsPH. In Section~\ref{sec_clouds} we consider the construction of stationary Proca clouds around Kerr and obtain one existence line for a particular set of ``quantum" numbers. In Section~\ref{sec_stars} we shall briefly review some of the main features of Proca stars, that form a limiting case of KBHsPH, and discuss some of their physical properties in the rotating case. In Section~\ref{sec_kbhsph} we finally construct KBHsPH, discussing the ansatz, boundary conditions and solving numerically the field equations. Then we exhibit the domain of existence and phase space of one family of solutions, we discuss the Proca energy spacetime distribution and some other physical features of these BHs. We close with a discussion of the results of this paper and some of the open directions for future related research.  In the Appendices we provide some technical results, including the explicit expression for the Einstein tensor, Proca energy-momentum tensor and Proca field equations.

\section{Einstein--complex-Proca model}
\label{sec_model}
The field equations for a massive vector field were introduced by A. Proca~\cite{Proca} in the 1930s. Much more recently, gravitating Proca fields have been discussed by various authors -- see $e.g.$~\cite{Rosen:1994rq,Obukhov:1999ed,Toussaint:1999zz}. Here, we shall consider  two real Proca fields, both with mass $\mu$, but our discussion can be easily generalized to an arbitrary even number of real Proca fields (or an arbitrary number of complex ones). 
The two fields are described by the potential 1-forms $A^{(i)}$, $i=1,2$, and field strengths $F^{(i)}=dA^{(i)}$. It is convenient to organize them into a single complex Proca field:
\begin{equation}
\mathcal{A}=A^{(1)}+iA^{(2)} \ , \qquad \mathcal{F}=F^{(1)}+iF^{(2)} \ .
\end{equation}
We denote the complex conjugate by an overbar,
\begin{equation}
\bar{\mathcal{A}}=A^{(1)}-iA^{(2)} \ , \qquad \bar{\mathcal{F}}=F^{(1)}-iF^{(2)} \ .
\end{equation}
Considering that the two Proca fields do not couple to each other and couple minimally to gravity, one obtains the minimal Einstein--complex-Proca model, which is  
described by the action:
\begin{equation}
\label{action}
\mathcal{S}=\int d^4x \sqrt{-g}\left(\frac{1}{16 \pi  G}R
-\frac{1}{4}\mathcal{F}_{\alpha\beta}\bar{\mathcal{F}}^{\alpha\beta}
-\frac{1}{2}\mu^2\mathcal{A}_\alpha\bar{\mathcal{A}}^\alpha\right) \ .
\end{equation}
This (or its version with $A_\alpha$ real) 
is the action considered by previous studies of the Einstein-Proca model, see
$e.g.$ refs. \cite{Rosen:1994rq,Vuille:2002qz}.\footnote{We remark that these works did not succeed in finding regular particle-like solutions or BHs with Proca hair.}

Varying  (\ref{action}) $w.r.t.$ the potential $A_\alpha$ yields the Proca field equations
\begin{equation}
\nabla_\alpha\mathcal{F}^{\alpha\beta}=\mu^2 \mathcal{A}^\beta \ .
\label{procafe}
\end{equation}
Observe that these equations \textit{completely} determine $ \mathcal{A}^\beta$ once $\mathcal{F}^{\alpha\beta}$ is known. Thus, the Proca potential is not subject to gauge transformations, unlike the Maxwell potential, and it is as physical as the field strength. In particular \eqref{procafe} imply the Lorentz condition, thus a dynamical requirement, rather than a gauge choice:
\begin{equation}
\nabla_\alpha\mathcal{A}^\alpha = 0 \ .
\label{lorentz}
\end{equation}

As usual, the Einstein equations are found by taking the variation of
(\ref{action}) $w.r.t.$ the metric tensor $g_{\alpha \beta}$
\begin{equation}
R_{\alpha \beta}-\frac{1}{2}R g_{\alpha \beta}=8 \pi G T_{\alpha \beta} \ ,
\label{Einstein-eqs}
\end{equation}
where the energy-momentum tensor reads:
\begin{eqnarray}
T_{\alpha\beta}=\frac{1}{2}
( \mathcal{F}_{\alpha \sigma }\bar{\mathcal{F}}_{\beta \gamma}
+\bar{\mathcal{F}}_{\alpha \sigma } \mathcal{F}_{\beta \gamma}
)g^{\sigma \gamma}
-\frac{1}{4}g_{\alpha\beta}\mathcal{F}_{\sigma\tau}\bar{\mathcal{F}}^{\sigma\tau}+\frac{1}{2}\mu^2\left[  
\mathcal{A}_{\alpha}\bar{\mathcal{A}}_{\beta}
+\bar{\mathcal{A}}_{\alpha}\mathcal{A}_{\beta}
-g_{\alpha\beta} \mathcal{A}_\sigma\bar{\mathcal{A}}^\sigma\right]\ . \ \ \ \ \ \ \  \ 
\label{procaemt}
\end{eqnarray}

The action possesses a global $U(1)$ symmetry, since it is invariant under the transformation 
$\mathcal{A}_\beta\rightarrow e^{i\chi}\mathcal{A}_\beta$, with $\chi$ constant; 
this implies the existence of a  4-current, 
\begin{equation}
\label{j}
j^\alpha=\frac{i}{2}\left[\bar{\mathcal{F}}^{\alpha \beta}\mathcal{A}_\beta-\mathcal{F}^{\alpha\beta}\bar{\mathcal{A}}_\beta\right] \ ,
\end{equation}
which is conserved by virtue of the field equations (\ref{procafe}): $\nabla_\alpha j^\alpha=0$. Consequently, there exists a Noether charge, $Q$, obtained integrating the temporal component of the 4-current on a space-like slice $\Sigma$:
\begin{equation}
Q=\int_\Sigma d^3x j^0 \ .
\label{q}
\end{equation}
We emphasize that unlike the massless limit of the theory, wherein the global symmetry becomes local, the last integral cannot be converted into a surface integral. In other words, there is no Gauss law.

\section{No Proca-hair theorems}
\label{sec_nohair}
If one considers Maxwell's equations for a test field with a spherically symmetric ansatz (a purely radial electric field) on the Schwarzschild background one finds a regular solution on and outside the Schwarzschild horizon ($cf.$ Section 2.1 in~\cite{Herdeiro:2015waa}). This is a smoking gun that a spherically symmetric field can be added, non-linearly, to the Schwarzschild solution, which indeed yields the well-known Reissner-Nordstr\"om BH. 
Adding a mass term to the Maxwell field -- hence converting it into a Proca field -- drastically alters the behaviour of the test field solution: it is not possible to find a solution which is both finite at the horizon and at spatial infinity, no matter how small $\mu$ is. 
In particular, for the asymptotically (exponentially) decaying solution, the Proca potential squared diverges at the horizon~\cite{Gottlieb:1984jg} - see Section~\ref{sec_nohair2}. Thus, requiring any amount of Proca field in equilibrium outside the horizon implies an infinite pile up of Proca invariants at the horizon. This behaviour parallels that of a scalar field (massless or massive) discussed in~\cite{Herdeiro:2015waa} and it is intimately connected with the existence/absence of a Gauss law for the Maxwell/Proca field. Moreover, it shows one cannot find a regular, spherically symmetric BH solution with Proca (time-independent) hair bifurcating from the Schwarzschild solution.

We shall review in~Section \ref{sec_nohair1} a more robust argument for the inexistence of stationary BHs with Proca hair, due to Bekenstein~\cite{Bekenstein:1971hc,Bekenstein:1972ky}, and that applies to our model~\eqref{action}. A fundamental assumption in the argument is that the Proca field and the background share the same symmetries. This \textit{symmetry inheritance} of the spacetime symmetries by the matter fields is precisely the assumption that  the KBHsPH presented later in this paper will violate. Then, in Section \ref{sec_nohair2}, we show that even dropping the symmetry inheritance assumption one can establish a no-hair theorem, for \textit{spherically symmetric} BHs. This is compatible with the KBHsPH solutions presented here, which are stationary and axi-symmetric, and shows that these solutions cannot have a static limit. This fact is in agreement with the domain of existence of KBHsPH, $cf.$ Section~\ref{subsec_III}.

\subsection{Bekenstein's theorem}
\label{sec_nohair1}
Following Bekenstein~\cite{Bekenstein:1971hc,Bekenstein:1972ky}, we consider a rotating, stationary, asymptotically flat BH spacetime. For matter obeying the null energy condition, the rigidity theorem implies that the spacetime is also axi-symmetric~\cite{Hawking:1971vc}. We write the spacetime metric in coordinates adapted to these symmetries $(t,r,\theta,\phi)$, so that the two Killing vector fields read ${\bf k}=\partial_t$, ${\bf m}=\partial_\phi$.

For simplicity we consider the Proca field to be real. But the proof generalizes straightforwardly for an arbitrary number of real Proca fields, and in particular for a complex Proca field. We denote the real Proca potential and field strength as $A_\alpha$ and $F_{\alpha\beta}$, respectively. We assume that this field inherits the spacetime symmetries. In particular for the coordinates chosen above 
this means that: 
\begin{equation}
\mathcal{L}_{\bf k} A_\alpha=\mathcal{L}_{\bf m} A_\alpha=0=\mathcal{L}_{\bf k} F_{\alpha\beta}=\mathcal{L}_{\bf m} F_{\alpha\beta} \ .
\label{symmetriesaf}
\end{equation}

The proof proceeds as follows. We contract the Proca equation $\nabla_\alpha F^{\alpha\beta}=\mu^2 A^\beta$ with $A_\beta$ and integrate over the BH exterior space-time:
\begin{equation}
\int d^4x\sqrt{-g}\left[A_\beta\nabla_\alpha F^{\alpha\beta}-\mu^2 A_\beta A^\beta\right]=0 \ .
\end{equation}
Next, integrating the first term by parts:
\begin{equation}
\int d^4x\sqrt{-g}\left[\frac{F_{\alpha\beta} F^{\alpha\beta}}{2}+\mu^2 A_\beta A^\beta\right]-\int_{\mathcal{H}}d^3\sigma n^\alpha A_\beta F_{\alpha}^{\ \beta}=0 \ ,
\label{bek1}
\end{equation}
where the boundary term is computed on the (spatial section of the) horizon, $\mathcal{H}$, and the other boundary term (at infinity) vanishes since the Proca field falls off exponentially fast.

Now we argue that the boundary term in  \eqref{bek1} is zero. To do so, we first observe that defining $b_\alpha\equiv A_\beta F_{\alpha}^{\ \beta}$, then $b_t=0=b_\phi$. This results from the symmetries imposed, which imply $A_r=A_\theta=F_{r\theta}=F_{t\phi}=0$.\footnote{$F_{t\phi}=0$ follows immediately from~\eqref{symmetriesaf}. Non-vanishing $A_r,A_\theta,F_{r\theta}$ would imply non vanishing components $T_{tr}$ and $T_{t\theta}$ of the energy momentum tensor~\eqref{procaemt}, which are incompatible with the symmetries of the problem.} 
Since, the event horizon of a stationary, asymptotically flat spacetime is a Killing horizon, the normal to $\mathcal{H}$, $n^\alpha$, 
is a linear combination of the Killing vector fields. Then $n^\alpha A_\beta F_{\alpha}^{\ \beta}=0$. 
We conclude that\footnote{We are implicitly assuming that $d^3\sigma$ and $A_\alpha,F_{\alpha\beta}$ are finite on $\mathcal{H}$. This assumption actually breaks down for the massless case (Maxwell field) due to gauge invariance.}
\begin{equation}
\int d^4x\sqrt{-g}\left[\frac{F_{\alpha\beta} F^{\alpha\beta}}{2}+\mu^2 A_\beta A^\beta\right]=0 \ .
\label{bek3}
\end{equation}

Contrary to the scalar field case (see $e.g.$~\cite{Herdeiro:2015waa}) this integrand is not positive definite. Thus, a further argument is necessary, which can be constructed by using an orthonormal basis, which we denote as $\{ {\bf e}^{\underline{t}},{\bf e}^{\underline{r}},{\bf e}^{\underline{\theta}},{\bf e}^{\underline{\phi}}\}$. Flat (underlined) indices are raised and lowered with the standard Cartesian Minkowski metric. Taking into account the allowed components by symmetry of the Proca potential and field strength, \eqref{bek3} becomes:
\begin{equation}
\begin{array}{l}
\displaystyle{\int d^4x\sqrt{-g}\left[(F_{\underline{t}\underline{r}})^2+(F_{\underline{t}\underline{\theta}})^2+(A_{\underline{t}})^2\right]} \displaystyle{=\int d^4x\sqrt{-g}\left[(F_{\underline{\phi}\underline{r}})^2+(F_{\underline{\phi}\underline{\theta}})^2+(A_{\underline{\phi}})^2\right]} \ .
\end{array}
\label{procaproof}
\end{equation}
Analysing the time-reversal invariance of the Proca equation, shows that $\{F_{\underline{t}\underline{r}},F_{\underline{t}\underline{\theta}},A_{\underline{t}}\}$ are even, whereas $\{F_{\underline{\phi}\underline{r}},F_{\underline{\phi}\underline{\theta}},A_{\underline{\phi}}\}$ are odd, under time-reversal. Thus, expanding the Proca potential and field strength in a power series of the angular momentum of the background, the first (second) set of field/potential components contains only even (odd) powers. The zeroth order terms only get contributions from the left hand side of \eqref{procaproof}; since the corresponding integrand is strictly positive and the integral is zero, the zeroth order terms must vanish. Then, the first order terms only get contributions from the right hand side of \eqref{procaproof}; since the corresponding integrand is strictly positive and the integral is zero, the first order terms must vanish. In this way one shows iteratively that the Proca field/potential must vanish, and hence there is no Proca hair. 
Observe that this theorem did not use the Einstein equations.

A different proof of the no Proca-hair theorem, possibly including a cosmological constant  and making use of the Einstein equations, 
has been given in \cite{Bhattacharya:2011dq}.

\subsection{A modified Pe\~{n}a--Sudarsky theorem}
\label{sec_nohair2}
The theorem of the previous subsection relied on the symmetry inheritance of the spacetime isometries by the Proca field. In particular the stationarity of the geometry implied a time-independence of the Proca potential/field. Recently, however, gravitating solitons composed by self-gravitating Proca fields were found by allowing the complex Proca field to have a \textit{harmonic time dependence}: Proca stars~\cite{Brito:2015pxa}. This time-dependence vanishes at the level of the energy momentum tensor and it is therefore compatible with a stationary geometry (see~\cite{Smolic:2015txa} for recent discussions of symmetry inheritance). Thus one may wonder if allowing the Proca field to have such harmonic time dependence allows for BHs with Proca hair. 

The situation just described parallels closely the well-known picture for complex scalar fields. The existence of scalar boson stars led Pe\~{n}a and Sudarsky to consider the possibility of spherically symmetric BH geometries with a scalar field possessing a harmonic time dependence. In this setup it was possible to establish a no-scalar-hair theorem, ruling out BHs with scalar hair even if the hair has such harmonic time-dependence~\cite{Pena:1997cy}. In the following we shall establish a no-Proca-hair theorem, allowing the complex Proca field to have a harmonic time dependence, for the case of spherical symmetry, by using a modified version of the arguments in~\cite{Pena:1997cy}. 

We consider a spherically symmetric line element with the parametrization (see $e.g.$~\cite{Brito:2015pxa}):
\begin{equation}
ds^2=-\sigma^2(r)N(r)dt^2+\frac{dr^2}{N(r)}+r^2d\Omega_2 \ ,  \qquad \ N(r)\equiv 1-\frac{2m(r)}{r} \ .
\label{ansatz1}
\end{equation}
The Ansatz we consider for the complex Proca potential is also the one introduced in \cite{Brito:2015pxa} for discussing spherical Proca stars and it is the most general one compatible with spherical symmetry and staticity:
\begin{equation}
\mathcal{A}=e^{-iwt}\left[f(r)dt+ig(r)dr \right] \ .
\label{ansatz2}
\end{equation}
In the above relations, $\sigma(r),m(r),f(r),g(r)$ 
are all real functions of the radial coordinate only and $w$ is the frequency parameter, which we take to
be positive without any loss of generality. 

 The Proca field equations~\eqref{procafe} yield 
 \begin{equation}
 \frac{d}{dr}\left\{\frac{r^2[f'(r)-wg(r)]}{\sigma(r)}\right\}=\frac{\mu^2r^2f(r)}{\sigma(r)N(r)} \ ,
 \label{proca1}
 \end{equation}
 and
 \begin{equation}
 f'(r)=wg(r) \left (1-\frac{\mu^2\sigma^2(r)N(r) }{w^2} \right) \ ,
 \label{eq1}
 \end{equation}
 where ``prime" denotes radial derivative. The Lorentz condition, \eqref{lorentz}, determines $f(r)$ in terms of the other functions:
\begin{equation}
f(r)=-\frac{\sigma(r)N(r)}{wr^2}\frac{d}{dr}\left[r^2\sigma(r)N(r)g(r)\right] \ ;
\label{lor}
\end{equation} 
this can be rewritten as
\begin{equation}
\frac{d}{dr}\left[r^2\sigma(r)N(r)g(r)\right] =-\frac{wr^2 f(r)}{\sigma(r)N(r)}\ .
\label{eq2}
\end{equation} 
 Observe that (\ref{proca1})-(\ref{eq1}) imply (\ref{eq2}), as they should. 
The essential Einstein equations, \eqref{Einstein-eqs}, read (there is a further Einstein equation which is a differential consequence of these)
\begin{eqnarray}
\label{Einstein-eqs1}
&&
m'=4\pi G r^2
\left[
\frac{(f'-wg)^2}{2\sigma^2}
+\frac{1}{2}\mu^2 \left(g^2N+\frac{f^2}{N\sigma^2}\right)
\right], \nonumber 
\\
&&\frac{\sigma'}{\sigma}=4\pi G r  \mu^2
\left(g^2+\frac{f^2}{N^2\sigma^2} \right).
\end{eqnarray} 
We also note that the $T_t^t$ component of the energy-momentum tensor -- the
energy density -- reads
 \begin{equation}
\label{ro}
-T^t_t=
 \frac{(f'-wg)^2}{2\sigma^2}
+\frac{1}{2}\mu^2 \left(g^2N+\frac{f^2}{N\sigma^2} \right)\ .
\end{equation}

To establish the no-Proca-hair theorem, let us assume the existence of a regular BH solution of the above  equations. Then the geometry would possess a non-extremal horizon at, say, 
$r=r_H>0$, which requires that 
 \begin{equation}
N(r_H)=0 \ ,
\end{equation}
since $r=r_H$ is a null surface. Since we are assuming that there are no more exterior horizons, then $r>r_H=$constant are timelike surfaces and $N'(r_H)>0$. Also, we can choose without loss of generality that $\sigma(r_H)>0$, since the equations of motion are invariant under $\sigma\rightarrow -\sigma$. It follows that $N(r)$ and $\sigma(r)$ are strictly positive functions for any $r>r_H$, as a consequence of the Einstein equations~\eqref{Einstein-eqs1} and the assumption that there are no further more exterior horizons.

The regularity of the horizon implies that the energy density of the Proca field
is finite there.
From (\ref{ro}) one can see that this implies
 \begin{equation}
f(r_H)=0\ .
\end{equation}
Then the function $f(r)$ starts from zero at the horizon 
and remains strictly positive (or negative) for some $r$-interval.
Now, let us assume $f'(r)>0$ for 
$r_H<r<r_1$. Thus $f(r)$, in this interval, is a strictly increasing (and positive) function 
(the case $f'(r)<0$ can  be discussed in a similar way).

Next, we consider the expression (which appears in~\eqref{eq1}) 
 \begin{equation}
P(r)\equiv 1-\frac{\mu^2\sigma^2(r)N(r) }{w^2}\ .
\end{equation}
One can see that $P(r_H)=1$; actually $P$  
becomes negative for large $r$, since $N\to 1$, $\sigma\to 1$ as $r\to \infty$, while $\mu>w$, which is a bound state condition necessary for an exponential decay of the Proca field at infinity. But the important point is the existence of an $r-$interval
$r_H<r<r_2$ where $P$ is a strictly positive function.

Let $r_c$ be the minimum between $r_1$ and $r_2$.
Then   
  we observe that (\ref{eq2}) implies
\begin{equation}
 r^2\sigma(r)N(r)g(r)  =-w\int_{r_H}^r\ dx \frac{x^2}{\sigma(x)N(x)} f(x)<0 
\label{eq21}
\end{equation} 
for  any $r$ in the interval $r_H<r<r_c$. Consequently, $g(r)<0$ in this interval, since $\sigma,N$ are positive everywhere outside the horizon.

The last conclusion implies a contradiction: $g(r)<0$ is not compatible with $f'(r)>0$, in that interval. In fact, $f'(r)>0$ together with $P>0$ and $w>0$, from (\ref{eq1}), that $g(r)>0$. 
Thus we conclude that $f(r)=g(r)=0$ is the only solution compatible with a BH geometry ($q.e.d.$).

One final observation concerning static fields ($w=0$). In such cases, one has only an electric potential, $g(r)=0$.
Then, the Proca equations  on a Schwarzschild
background -- $i.e.$ taking the line element (\ref{ansatz1}) 
with $\sigma(r)=1$, $N(r)=1-2M/r$, --
can be solved in closed form by taking the ansatz~\cite{Gottlieb:1984jg}
\begin{equation}
f(r)=\frac{e^{-\mu r}}{r}S(r) \ ,
\label{eqs1}
\end{equation} 
where $S(r)$ is a solution of the Kummer equation~\cite{Abramowitz} 
\begin{equation}
z\frac{d^2 S(z)}{dz^2}-z \frac{d S(z)}{dz}-M \mu S(z)=0\ ,~~{\rm with}~~z\equiv 2\mu(r-2M) \ .
\label{eqs2}
\end{equation} 
This equation possesses a solution which is regular on and outside the  horizon.
In particular, $S(z)$ takes a constant nonzero value at $z=0$ ($i.e.$ $r=2M$).
This implies, however, that the invariant $A_\mu A^\mu=-f^2(r)/(1-2M/r)$
diverges at the horizon.


\section{Stationary Proca clouds around Kerr}
\label{sec_clouds}
The theorem of subsection~\ref{sec_nohair2} leaves open the possibility that \textit{stationary} (rather than static and spherically symmetric) BHs with Proca hair, possessing a harmonic time dependence, may exist. There is, moreover, a new physical ingredient in the stationary case which, indeed, makes their existence not only possible, but also natural: \textit{superradiance}. 

Sufficiently low frequency modes of a test Proca field, are amplified when scattering off a co-rotating Kerr BH, by extracting rotational energy and angular momentum from the BH, in a purely classical process. This process was studied in the slow rotation limit of Kerr in~\cite{Pani:2012vp,Pani:2012bp}, where it was used for placing bounds on the photon mass. Sufficiently high frequency modes, on the other hand (or any non-co-rotating mode), are partly absorbed in a similar scattering. 

The same two behaviours occur for gravitationally bound modes, with frequency lower than the Proca mass. These modes are generically \textit{quasi-bound} states, $i.e$ they have a complex frequency. Then, the amplified modes become an instability of the background. Moreover, at the threshold between the two behaviours (growing and decaying modes), one finds bound states with a real frequency, which we dub \textit{stationary Proca clouds around Kerr BHs}. We shall now sketch the study of the Proca bound states around Kerr BHs in a way suitable for the computation of KBHsPH. A more detailed account of stationary Proca clouds will appear elsewhere. 

We use the parametrization of the Kerr metric  introduced in~\cite{Herdeiro:2015gia}: 
\begin{eqnarray}
 ds^2=-e^{2F_0}N dt^2+e^{2F_1}\left(\frac{dr^2}{N}+r^2 d\theta^2\right) + e^{2F_2}r^2 \sin^2\theta \left(d\varphi-W dt\right)^2 \ ,
 \label{kerrnc}
\end{eqnarray}
where 
\begin{equation}
N\equiv 1 -\frac{r_H}{r} \ ,
\label{n}
\end{equation} 
and  $F_i,W$ are functions of the spheroidal coordinates $(r,\theta)$, which read, explicitly,\footnote{The parameters $b$ (here) and $c_t$ (in~\cite{Herdeiro:2015gia}) relate as $b=-c_t$.}
\begin{eqnarray}
\nonumber
&&
e^{2F_1}=\left(1+\frac{b}{r}\right)^2+b(b+r_H)\frac{\cos^2\theta}{r^2}\ ,
\\
\nonumber
&&
e^{2F_2}=e^{-2F_1}
\left\{
          \left[
\left(1+\frac{b}{r}\right)^2+\frac{b(b+r_H)}{r^2}
          \right]^2-b(b+r_H)\left(1-\frac{r_H}{r}\right)\frac{\sin^2\theta}{r^2}
\right\}\ ,
\\
\nonumber
&&
F_0=-F_2 \ , \\
&&
\label{functionsKerr}
\displaystyle{W=e^{-2(F_1+F_2)}
\sqrt{b(b+r_H)}(r_H+2b)
\frac{(1+\frac{b}{r})}{r^3}} \ .
\end{eqnarray}
The relation between these coordinates $(t,r,\theta,\varphi)$ and the standard Boyer-Lindquist coordinates $(t,R,\theta,\varphi)$ is simply a radial shift:
\begin{equation}
r=R-\frac{a^2}{R_H} \ , 
\end{equation}
where $R_H$ is the event horizon Boyer-Lindquist radial coordinate,  $R_H\equiv M+\sqrt{M^2-a^2}$, for a Kerr BH with mass $M$ and angular momentum $J=aM$. In the new coordinate system $(t,r,\theta,\varphi)$, the Kerr solution is parameterized by $r_H$ and $b$, which relate to the Boyer-Lindquist parameters as
\begin{equation}
r_H=R_H-\frac{a^2}{R_H} \ , \qquad b=\frac{a^2}{R_H} \ .
\end{equation}
Clearly, $r_H$ fixes the event horizon radius; $b$ is a spheroidal prolateness parameter (see Appendix \ref{prolatecoordinates}), and can be taken as a measure of non-staticity, since $b=0$ yields the Schwarzschild limit.

The ADM mass, ADM angular momentum and horizon angular velocity read, in terms of the parameters $r_H,b$ (we set $G=1=c$)
\begin{equation}
\begin{array}{c}
M=\frac{1}{2}(r_H+2b) \ , \ \ \ 
J=\frac{1}{2}\sqrt{b(b+r_H)}(r_H+2b) \ , \ \ \
\label{Kerr2}
\displaystyle{\Omega_H=\frac{1}{r_H+2b}\sqrt{\frac{b}{r_H+b}}} \ .
\end{array}
\end{equation}
The choice $r_H=-2b\neq 0$
yields Minkowski spacetime expressed in spheroidal prolate coordinates (Appendix \ref{prolatecoordinates}). Extremality occurs when $r_H=0$.

One considers the Proca field equations \eqref{procafe} on the background~\eqref{kerrnc}, using an ansatz given in terms of four functions $(H_i,V)$, all of which depends on $r,\theta$, and with a harmonic time and azimuthal dependence, which introduce a (positive) frequency, $w>0$, and the azimuthal harmonic index, $m\in \mathbb{Z}$:\footnote{We recall that in the scalar field case, the anstaz was  
 $\Psi=e^{i(m\varphi-wt)}R(r)S(\theta)$ 
for the stationary scalar clouds~\cite{Hod:2012px,Benone:2014ssa} and 
 $\Psi=e^{i(m\varphi-wt)}\phi(r,\theta)$ 
for the fully non-linear solutions~\cite{Herdeiro:2014goa}. In this case the test field analysis admits separation of variables, which does not occur for the Proca case.}
\begin{equation}
\mathcal{A}=e^{i(m\varphi-w t)}\left(
 iV dt  +H_1dr+H_2d\theta+i H_3 \sin \theta d\varphi 
\right) \ .
\label{procaclouds}
\end{equation}

Here we shall only address the case with $m=1$. The corresponding Proca equations are given in Appendix~\ref{appendixb}. These equations are solved with the following set of boundary conditions:
\begin{description}
\item[i)] at infinity, 
\begin{eqnarray}
  H_i|_{r=\infty}=V|_{r=\infty}=0\ ;
  \label{bccloudslarge}
\end{eqnarray}
\item[ii)] on the symmetry axis, 
\begin{equation}
H_1|_{\theta=0,\pi}
 = \partial_\theta H_2\big|_{\theta=0,\pi}=\partial_\theta H_3\big|_{\theta=0,\pi}=V|_{\theta=0,\pi}=0 \ ;
 \label{bccloudsaxis}
\end{equation}
\item[iii)]Êat the event horizon $(r=r_H)$ the boundary conditions become simpler by introducing a new radial coordinate $x\equiv \sqrt{r^2-r_H^2}$, such that the horizon is located at $x=0$.
Then one imposes
\begin{eqnarray}
 H_1|_{x=0}=\partial_x H_2|_{x=0}=\partial_x H_3|_{x=0}= 0 \ , \qquad  \left(V+\frac{w}{m}H_3\sin\theta\right)|_{x=0}=0  
\label{bccloudshorizon} \ . 
\end{eqnarray}
\end{description}
These boundary conditions are compatible with an approximate construction of the $m=1$ solutions
on the boundary of the domain of integration. 
All such solutions we have constructed so far are symmetric $w.r.t.$ a reflection along the equatorial plane. This symmetry is imposed by taking
 \begin{equation}
 \partial_\theta H_1|_{\theta=\pi/2}=
  H_2\big|_{\theta=\pi/2}=\partial_\theta H_3\big|_{\theta=\pi/2}= \partial_\theta  V|_{\theta=\pi/2}=0\ .
\end{equation}
We remark, however, that odd-parity composite configurations are also likely to exist.
Moreover, we observe that for $m>1$ the boundary conditions satisfied 
by some of the gauge potentials
at $\theta=0,\pi$ are different.

We have solved the equations for $H_i,V$, with the above boundary  conditions, for
a fixed Kerr BH background, by using the numerical approach described $e.g.$ in \cite{Herdeiro:2014pka}
for non-linear stationary scalar clouds.
The input parameters are $w,m$ for the Proca functions and $r_H,b$ for the geometry.
Regularity of the Proca fields at the horizon imposes the \textit{synchronization condition} (see the discussions in~\cite{Benone:2014ssa,Brihaye:2014nba})
 \begin{eqnarray}
w=m\Omega_H~,
\label{synchronization} 
\end{eqnarray}
which precisely means the scalar clouds are modes at the threshold of the superradiant instability (unstable modes obey $w<m\Omega_H$).  Observe that with~\eqref{synchronization}, the last condition in~\eqref{bccloudshorizon} becomes
\begin{equation}
\label{condA}
\xi^\alpha \mathcal{A}_\alpha\big|_{r_H}=0 \ ,
\end{equation}
where $\xi^\alpha\partial_\alpha=\partial_t+\Omega_H\partial_\varphi$ is the event horizon null generator.\footnote{Observe that for a \textit{massless} vector field, $i.e.$ a Maxwell field, $\xi^\alpha \mathcal{A}_\alpha\big|_{r_H}$ corresponds 
to $\Phi_H$ --the co-rotating electric potential on the horizon, which is non-zero in a gauge where the gauge potential vanishes asymptotically~\cite{Townsend:1997ku}.}
Observe also that  $\mathcal{A}$ is preserved by the action of $\xi$: $\mathcal{L}_\xi\mathcal{A}_\alpha=0$. 
This is analogous to what occurs in the scalar case ($\mathcal{L}_\xi\Psi=0$), but it is in contrast to the assumptions of Bekenstein's theorem, where it is required that the components of the Proca potential are invariant under $\partial_t$ and $\partial_\varphi$ separately, $cf.$~\eqref{symmetriesaf}.

\begin{figure}[h!]
  \begin{center}
    \includegraphics[width=9.0cm]{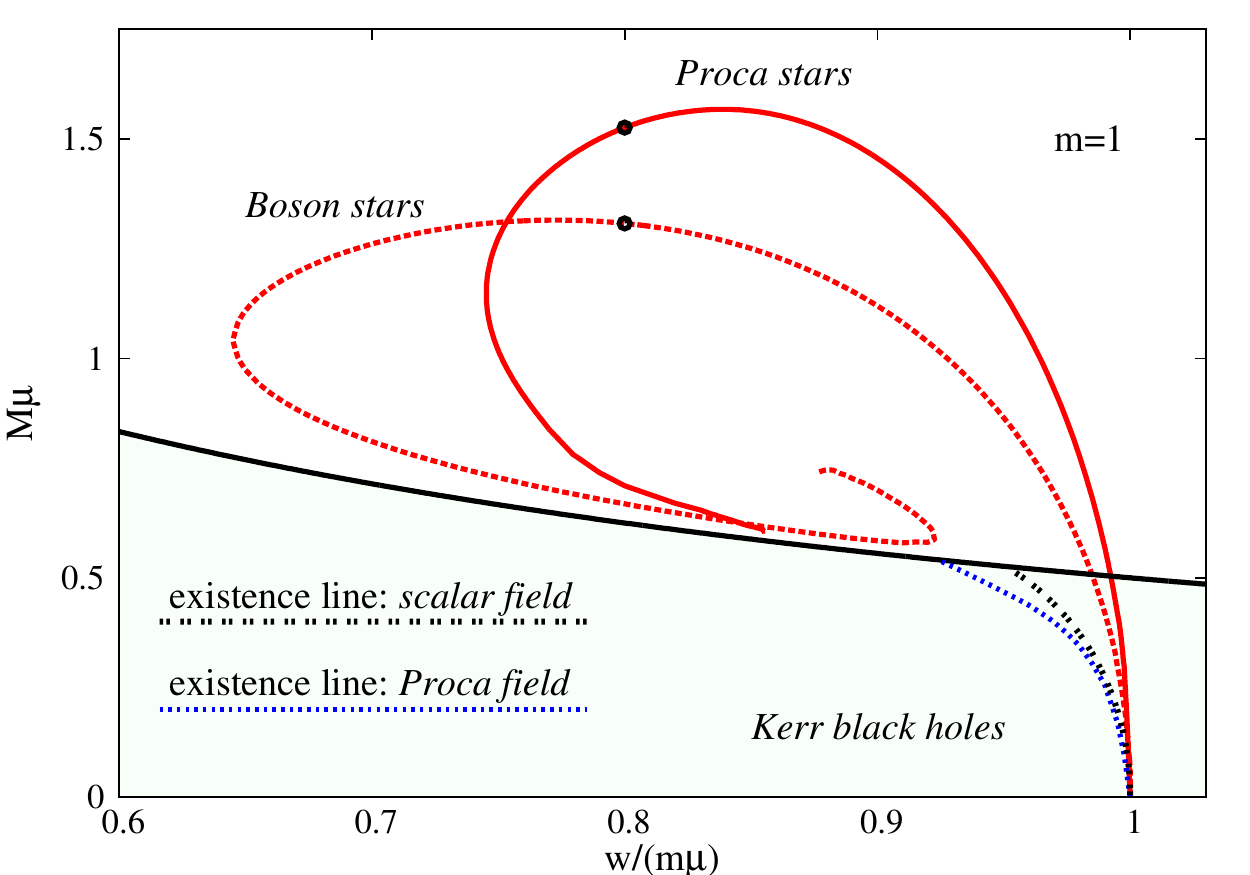}
  \end{center}
 \caption{Existence line for Proca stationary clouds with $m=1$ (blue dotted line) and the comparable existence line for scalar stationary clouds (with $m=1=\ell$, $n=0$, $cf.$~\cite{Benone:2014ssa}, black double dotted line), in an ADM mass $vs.$ frequency $w/m=\Omega_H$ diagram of Kerr BHs. Both axes are shown in units of the scalar/Proca field mass $\mu$. The black solid line corresponds to extremal Kerr BHs and non-extremal solutions exist below that line. Two red lines describing scalar boson stars (dotted) and Proca stars (solid) are also shown, that will be described in the next section.}
  \label{clouds}
\end{figure}

For a fixed $m$ ($m=1$ for the case here), for a given $w$ in some interval $w_{min}<w<\mu$, one finds a solution,  $i.e.$ the numerical iteration converges,  for a single value of $r_H$. Since $\Omega_H$ is determined by~\eqref{synchronization}, the corresponding mass is determined by~\eqref{Kerr2}. In other words, the regularity of the bound state implies a quantization condition of the background parameters; for each $m$, there is an \textit{existence line} in a $(M,\Omega_H)$ diagram representation of Kerr BHs, corresponding to a 1-dimensional subspace of the 2-dimensional Kerr parameter space. In Fig.~\ref{clouds} we exhibit the $m=1$ existence line (blue dotted line), which forms one of the boundaries of the domain of existence of KBHsPH. As we shall see in Section~\ref{subsec_III}, this line is one of the boundaries of the domain of existence of KBHsPH, which demonstrates that in the limit of small Proca field, KBHsPH reduce to the Kerr solutions that can support stationary Proca clouds, and hence that they are the non-linear realization of the clouds we have just discussed. 

It is interesting to compare the location of the existence lines for the Proca and scalar case in the Kerr $(M,\Omega_H)$ diagram, Fig.~\ref{clouds}. Comparing the $m=1$ existence line for stationary Proca clouds with the $m=\ell=1$ existence line for stationary scalar clouds,\footnote{The stationary scalar clouds are labelled by 3 quantum numbers $(\ell,m,n)$. The $n=0$, $\ell=m=1$ line is the one with smaller values of $\Omega_H$ for fixed $M$ for all possible values of $\ell,n$ and $m=1$~\cite{Benone:2014ssa}.} one observes that the former has smaller values of $\Omega_H$ for the same mass. This means, in particular, that there are Kerr BHs that are superradiantly stable against all $m=1$ scalar perturbations but are superradiantly unstable against $m=1$ Proca modes. A similar feature has been observed comparing the existence lines for \textit{Maxwell} and scalar stationary clouds in Kerr-AdS~\cite{Wang:2015fgp}. Finally let us remark that it was observed in~\cite{Herdeiro:2014pka} that including certain classes of self-interactions in the scalar field model, stationary scalar clouds can exist in an open set of the $(M,\Omega_H)$, rather than just a 1-dimensional line.  It is likely a similar result applies to self-interacting Proca fields, in view of the results in~\cite{Loginov:2015rya}.

We close this section by commenting on the node number of these stationary Proca clouds. 
In the scalar case, the number of nodes $n$ of the radial function defining the scalar field profile, 
is $n=0$ for fundamental states and $n\in \mathbb{N}$ for excited states. 
This issue becomes more subtle for Proca clouds (and Proca stars), 
since one has more than one potential component. Nevertheless, we remark that the all states we have obtained so far have always (only)
one node for the temporal component of the Proca potential  $V$, and thus are likely to represent the fundamental
modes of the problem.\footnote{The electric potential of the $m=0$ 
spherically symmetric Proca stars necessarily possesses at least one node \cite{Brito:2015pxa}.
Although the proof there cannot be generalized to the axially symmetric case,
we could not find any numerical indication for the existence of $m\geq 1$ nodeless solutions. 
}

\section{Spinning Proca stars} 
\label{sec_stars}
The stationary Proca clouds described in the previous section form one of the central ingredients to understand KBHsPH. They also form a part of the boundary of the domain of existence of these BHs, as we shall see in the next section. The other central ingredient corresponds to Proca stars, which again will form a part of the boundary of the domain of existence of KBHsPH. We shall now briefly review the relevant properties of these solutions, recently found in~\cite{Brito:2015pxa}, for understanding KBHsPH.

Proca stars can be either spherically symmetric and static or axially symmetric and stationary. The former are found by taking the ansatz~\eqref{ansatz1} for the line element and~\eqref{ansatz2} for the Proca field. With this ansatz, however, there are no BH solutions as shown in subsection~\ref{sec_nohair2}. The latter are found by taking a metric ansatz of the form~\eqref{kerrnc}, with $r_H=0$, with unspecified functions $F_0,F_1,F_2$ and the Proca potential ansatz~\eqref{procaclouds}, with unspecified functions $V,H_2,H_3$.  The remaining two (unspecified) functions are replaced as
\begin{equation}
W\rightarrow \frac{{W}}{r} \ , \qquad H_1\rightarrow \frac{{H_1}}{r} \ . 
\label{ww}
\end{equation}
We find it preferable to work with the new ${W},{H_1}$ when dealing with stars, due to their boundary conditions at the origin (rather than at a horizon). In the remaining of this section we shall always refer to these new functions. Solving the corresponding field equations with the following boundary conditions:
\begin{description}
\item[i)] at infinity,~\eqref{bccloudslarge}, together with
\begin{equation}
F_i\big|_{r=\infty}={W}\big|_{r=\infty}=0 \ , 
\label{bcstarslarge}
\end{equation}
\item[ii)] on the symmetry axis,~\eqref{bccloudsaxis}, together with 
\begin{equation}
 \partial_\theta F_i\big|_{\theta=0,\pi}=\partial_\theta {W}\big|_{\theta=0,\pi}=0
 \label{bcstarsaxis}
 \end{equation}
 \item[iii)] at the origin, 
 \begin{equation}
 \partial_r F_i\big|_{r=0}={W}\big|_{r=0}=H_i|_{r=0}=V|_{r=0}=0 \ .
 \end{equation}
\end{description} 
 Then, one finds a countable number of families of rotating Proca stars, labelled by $m\in \mathbb{Z}$, of which the cases with $m=1,2,3$ were discussed in~\cite{Brito:2015pxa}.  Therein, it was also found that, as for the scalar rotating boson stars, the ADM angular momentum and the Noether charge obey the simple relation 
 \begin{equation}
 J=mQ \ .
\label{amnc}
 \end{equation}
In Appendix~\ref{appendixc} we give a detailed derivation and discussion of this relation, which is more subtle in the case of Proca stars than for scalar boson stars.  Thus, following~\cite{Herdeiro:2014goa}, we define the normalized Noether charge, $q$, as 
 \begin{equation}
 q\equiv \frac{mQ}{J} \ ,
 \label{jq}
 \end{equation}
 which is obviously $q=1$ for all Proca stars, but will be $q\in [0,1]$ for KBHsPH.
 
 For $m=1$, the case in which we focus here, the Proca star solutions 
appear to form a spiral in an ADM mass, $M$, $vs.$ Proca field frequency, $w$, diagram, starting from $M=0$ for $w=\mu$, in which limit the Proca field becomes very diluted and the solution trivializes. 
At some intermediate frequency, a maximal ADM mass is attained. For $m=1$ this frequency is $w_{\rm max}/\mu=0.839$ and the maximal mass is $\mu M_{\rm max}=1.568$, a slightly larger value than for the corresponding scalar rotating boson star (for which $\mu M_{\rm max}=1.315$)~\cite{Brito:2015pxa}. 
 
 In Fig.~\ref{clouds}, we display the $m=1$ Proca star and scalar boson star curves (red solid and dotted lines). Comparing them, we observe: $(i)$ the slightly larger maximal mass for the Proca stars; $(ii)$ that the backbending of the inspiraling curve occurs, for Proca stars, for a larger value of the frequency parameter, and hence they exist in a narrower frequency interval; $(iii)$ that whereas for scalar boson stars with $m=1$ it was possible to obtain a third branch of solutions (after the second backbending) numerics become very difficult for Proca stars already on the second branch;\footnote{In the spherically symmetric case, 
the results in \cite{Brito:2015pxa} show the existence of a very similar picture for both 
  Proca stars and scalar boson stars, with the occurance  of secondary branches (together with the corresponding
spiral in a $(w,M)$-diagram) also in the former case.} for example, 
the function $F_0$ takes very large, negative values.
Finally, in complete analogy with the scalar boson star case, the Proca star line yields the second boundary of the domain of existence of KBHsPH; the latter reduce to Proca stars when the horizon size vanishes, as will be seen in the next section. 

\begin{figure}[h!]
  \begin{center}
    \includegraphics[width=8.1cm]{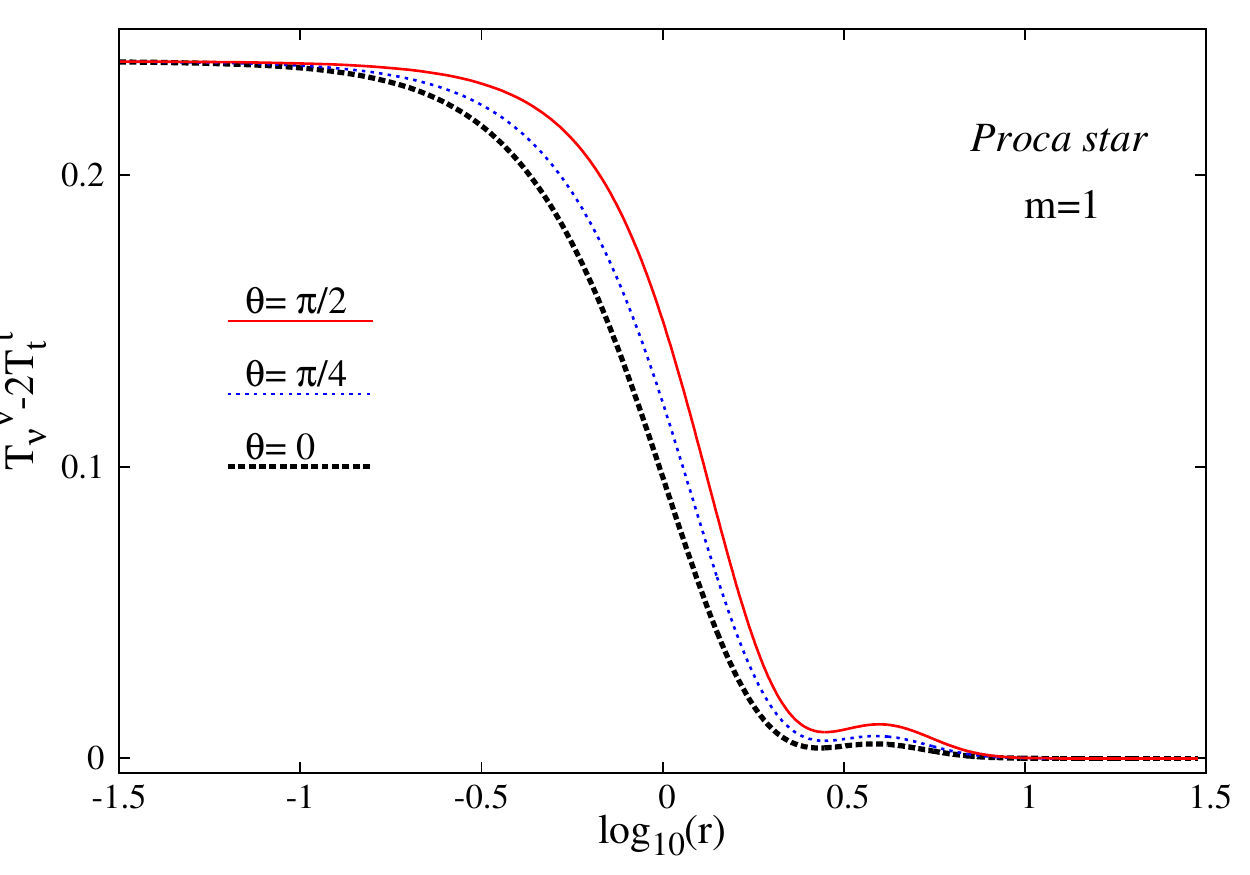}
       \includegraphics[width=8.1cm]{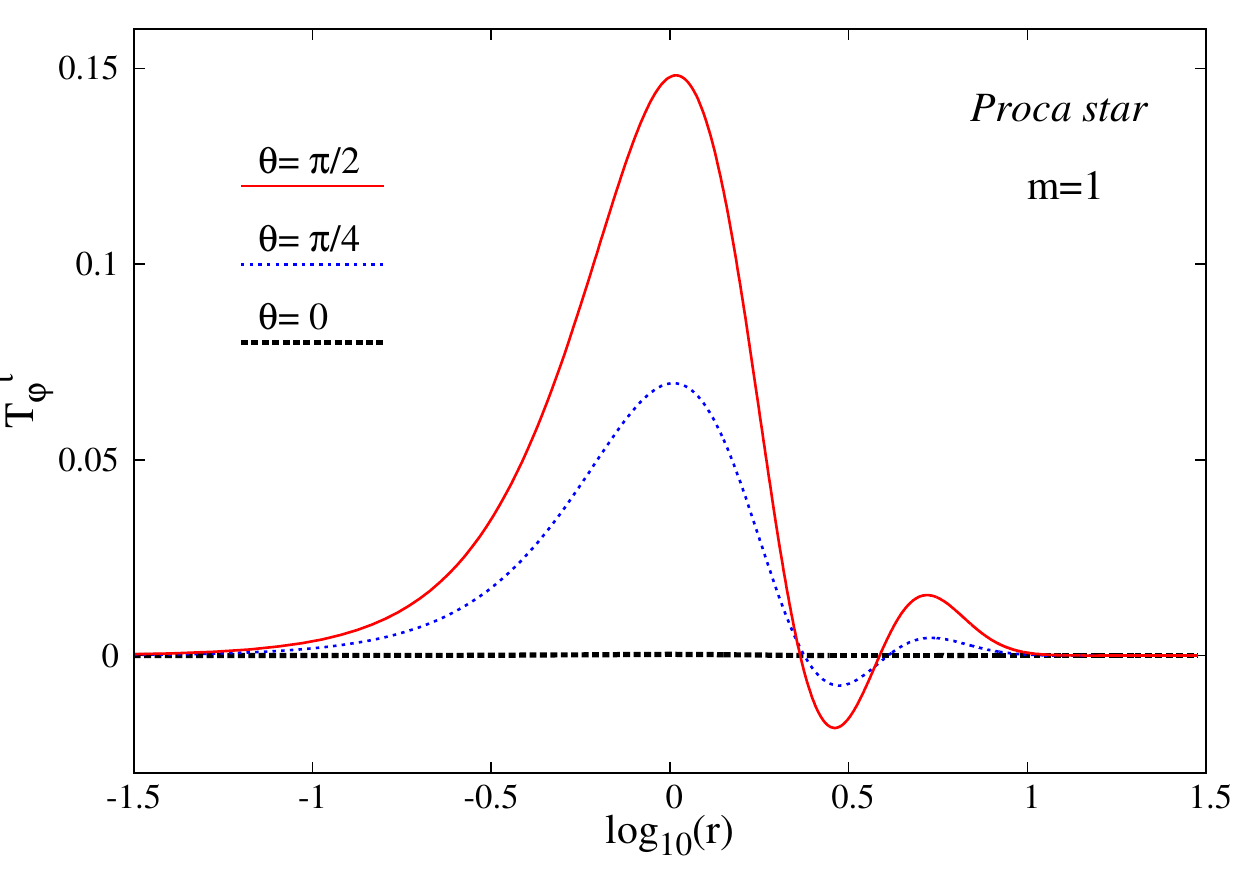}   
        \includegraphics[width=8.1cm]{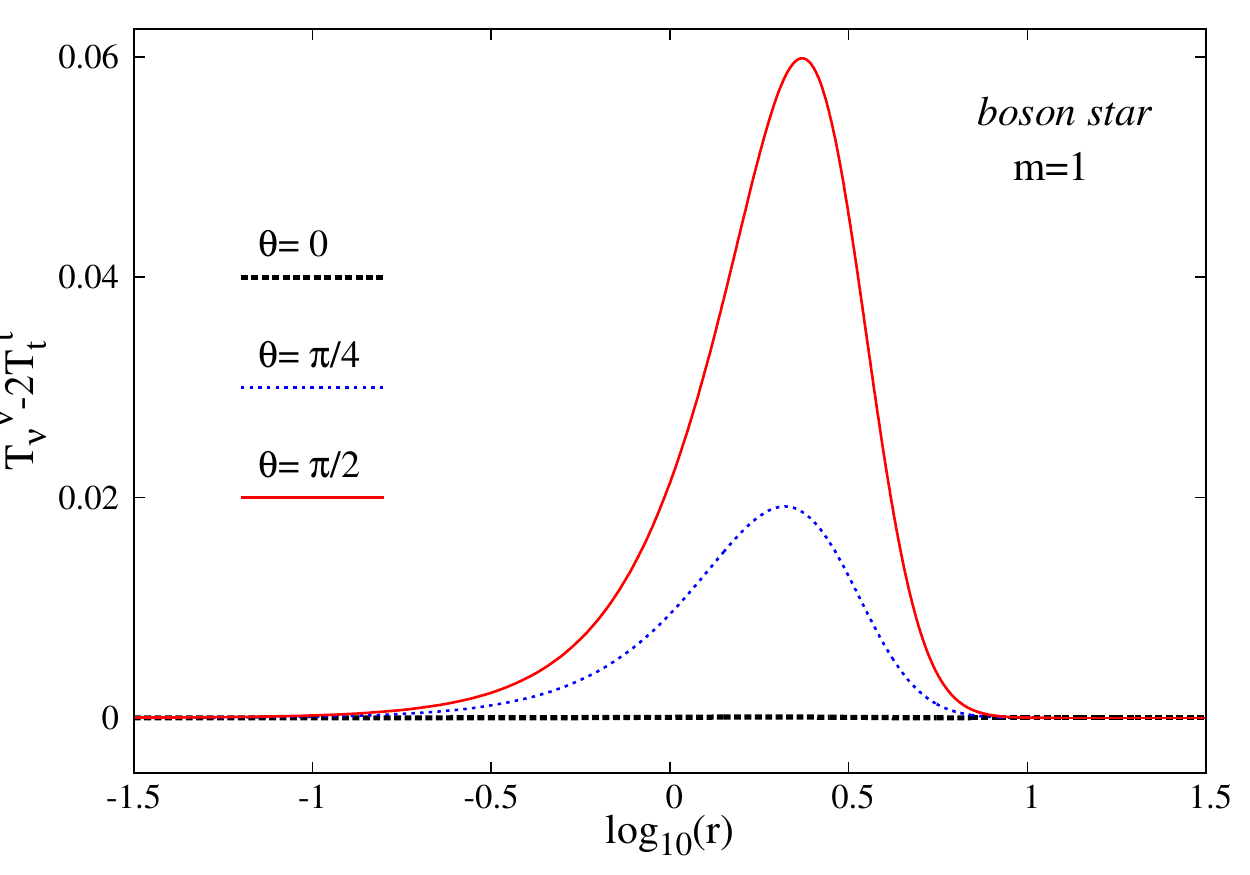}
      \includegraphics[width=8.1cm]{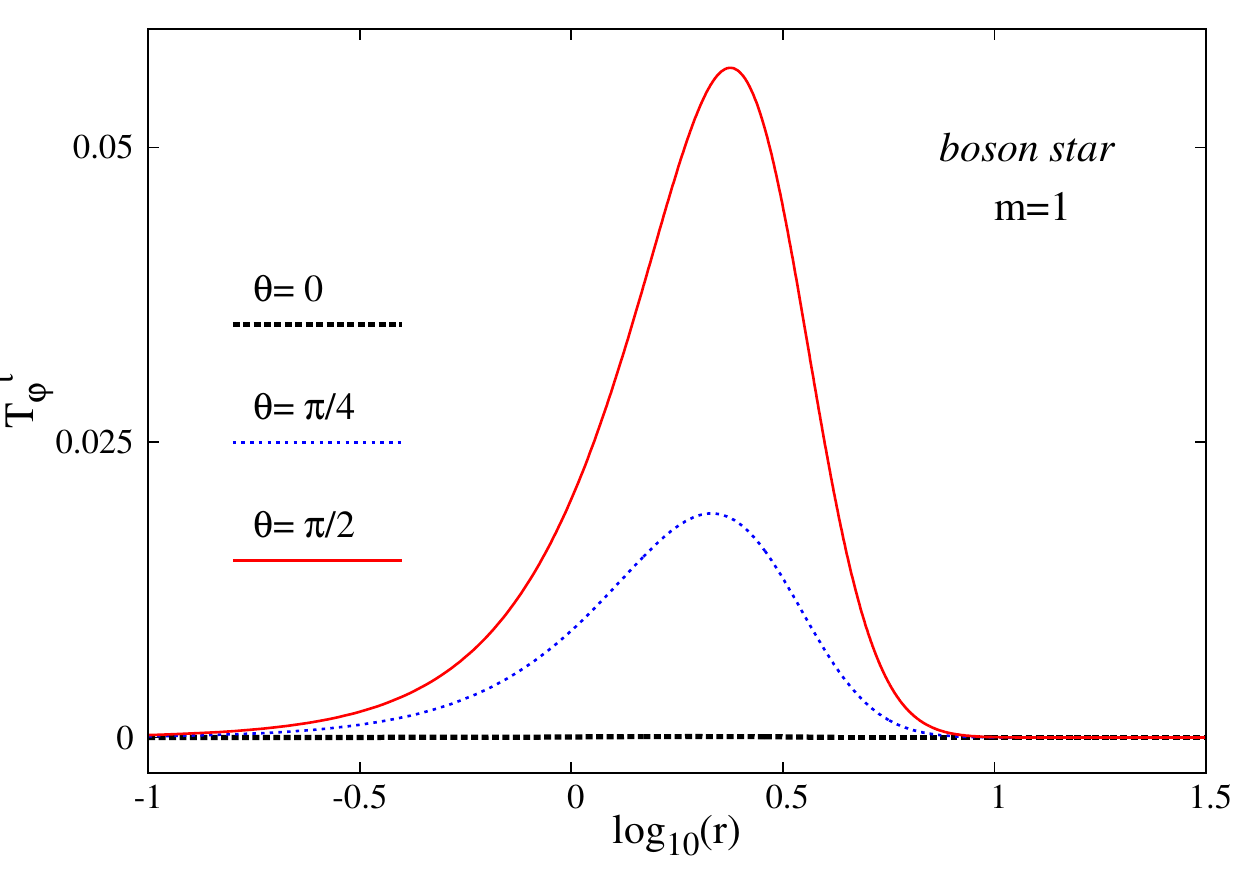}
  \end{center}
 \caption{Radial variation of the energy density, $cf.$~\eqref{ed} (left panel), and angular momentum density, $cf.$~\eqref{amd}  (right panel),  of the Proca field, for different constant $\theta$ sections of a spinning Proca star with $m=1$ (top panels) and a spinning scalar boson star with $m=1$ (bottom panels). Both solutions have $w=0.8$ and are marked with a bullet in Fig.~\ref{clouds}. The Proca star has $\mu M= 1.526$,  $\mu^2J= 1.575$, while the scalar boson star has $\mu M=1.308$, $\mu^2J=1.372$.}
  \label{PS1}
\end{figure}

\begin{figure}[h!]
  \begin{center}
    \includegraphics[width=8.1cm]{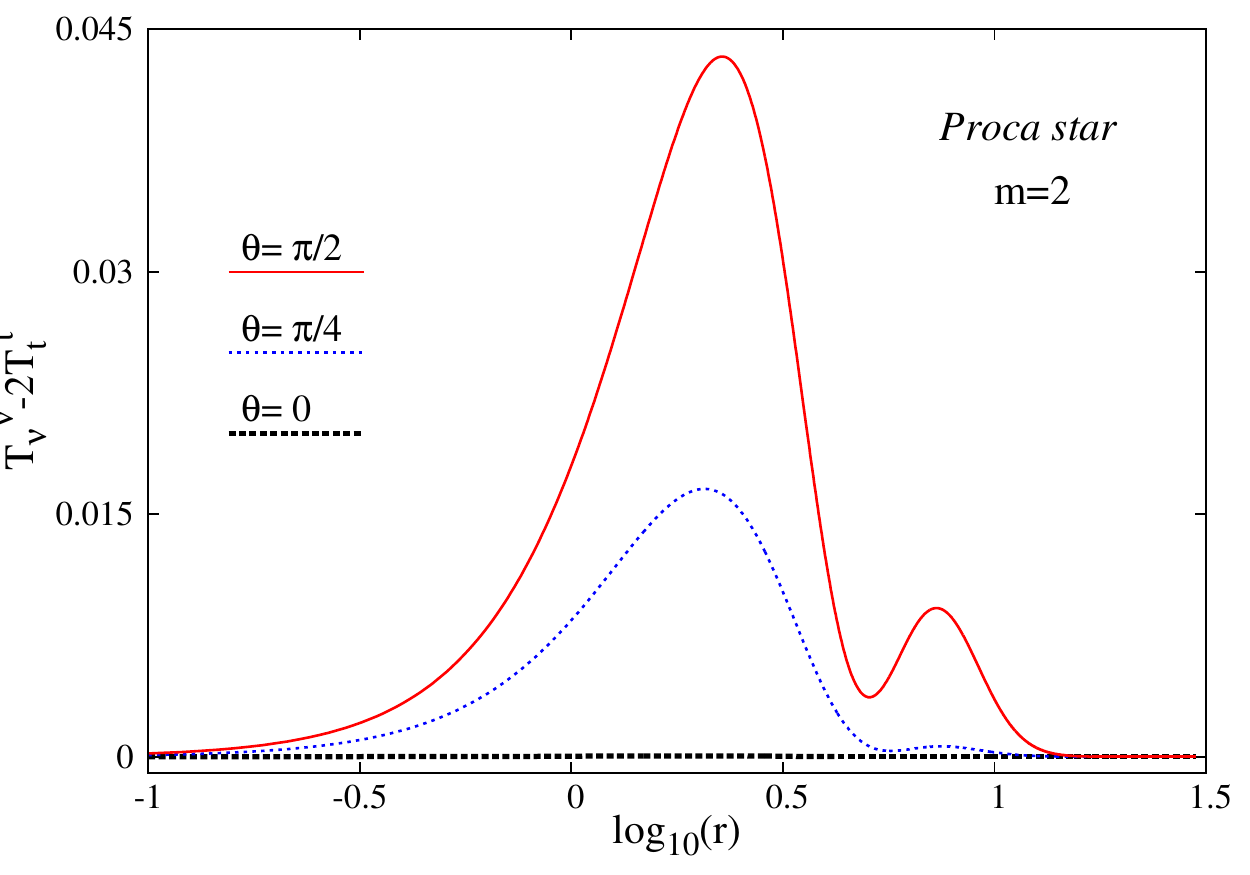}
       \includegraphics[width=8.1cm]{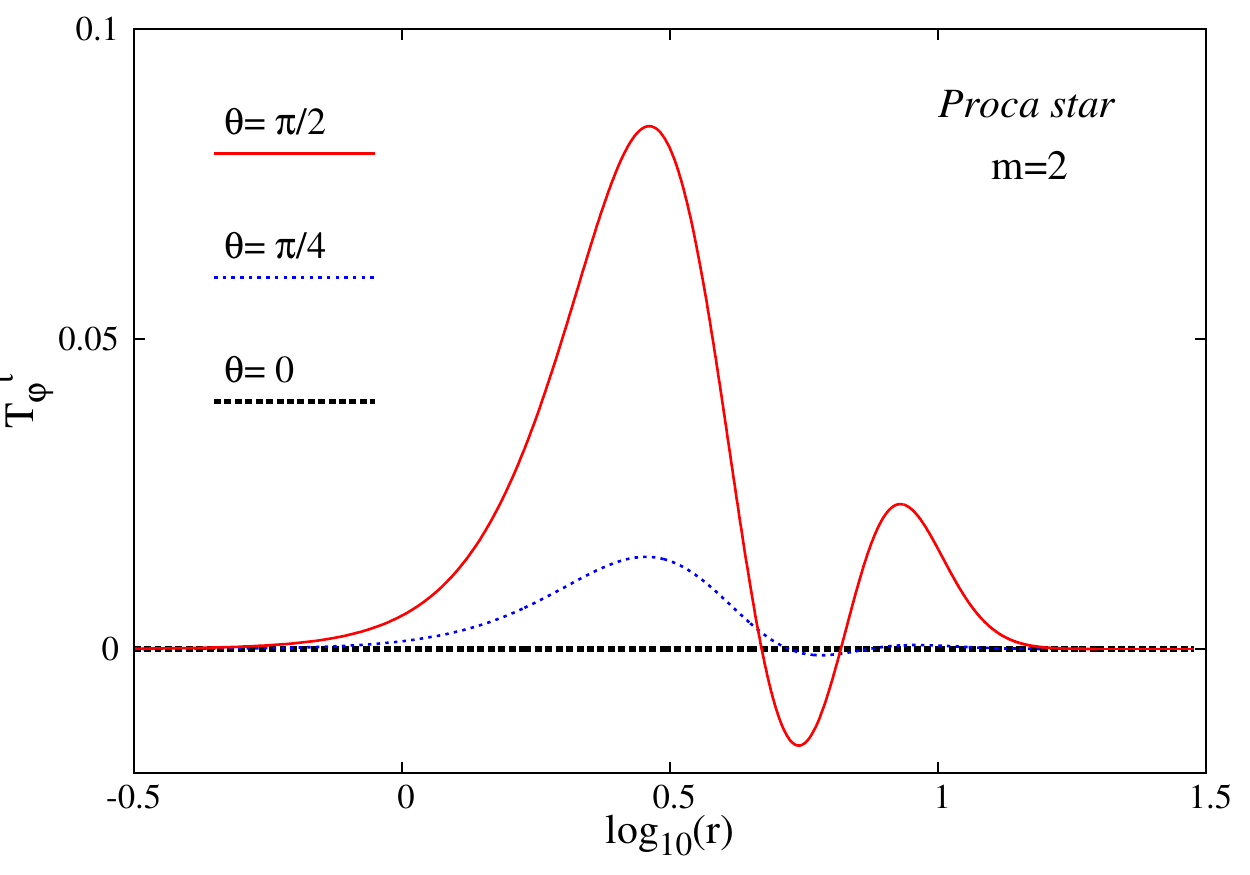}   
        \includegraphics[width=8.1cm]{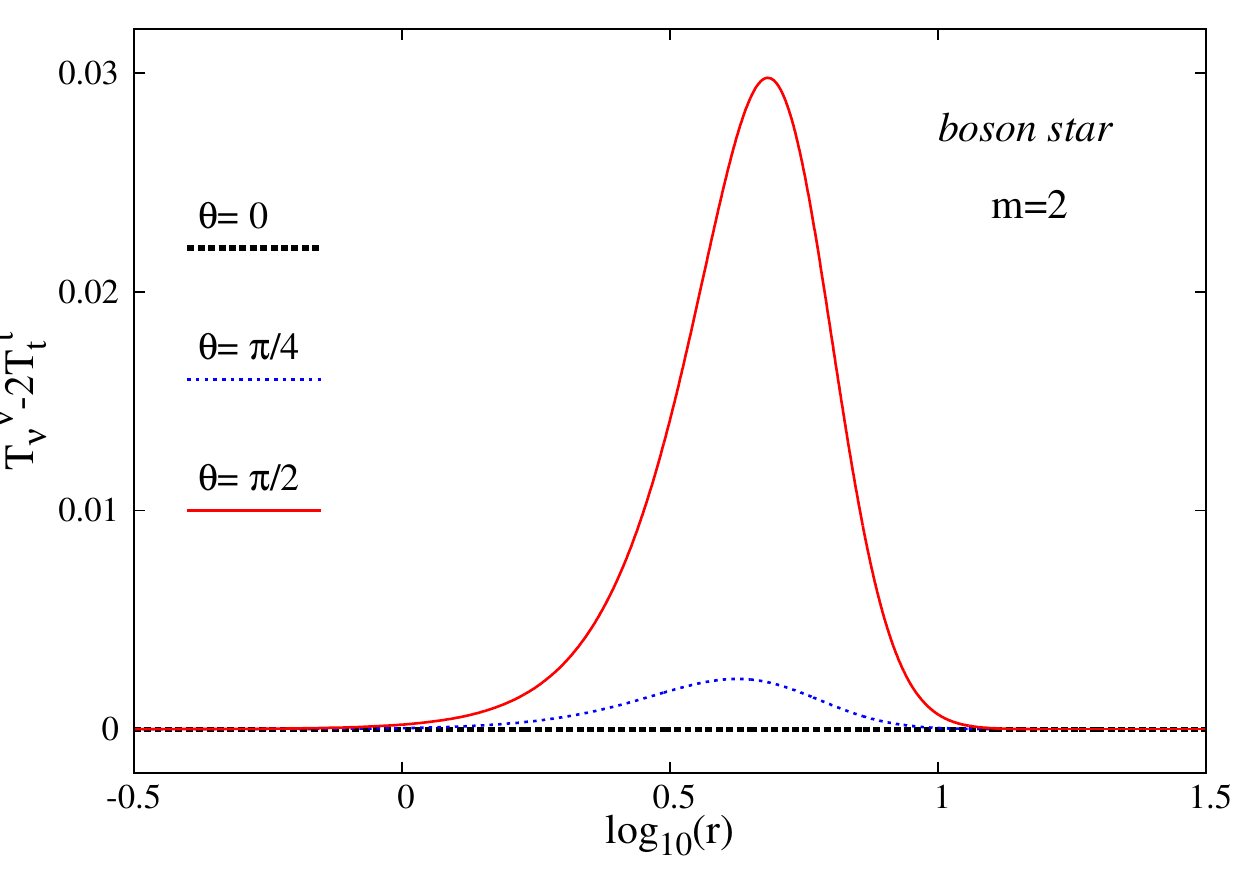}
      \includegraphics[width=8.1cm]{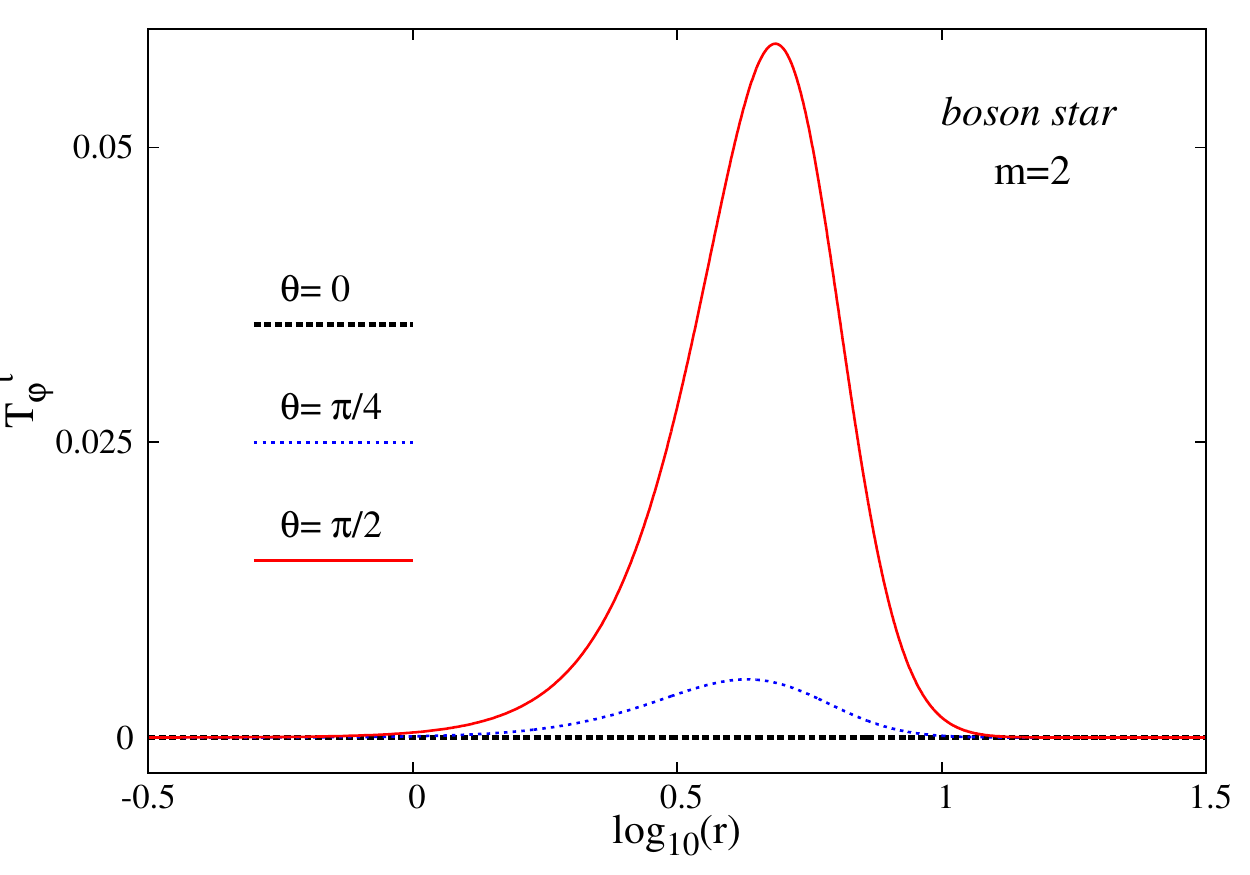}
  \end{center}
 \caption{Same as in Fig.~\ref{PS1} but for $m=2$. Both solutions have $w=0.8$. The Proca star has $\mu M=2.319$, 
$\mu^2J=4.873$ whereas the scalar boson star has $\mu M=2.016$, $\mu^2J=4.272$.}
  \label{PS2}
\end{figure}

\begin{figure}[h!]
  \begin{center}
    \includegraphics[width=6.1cm]{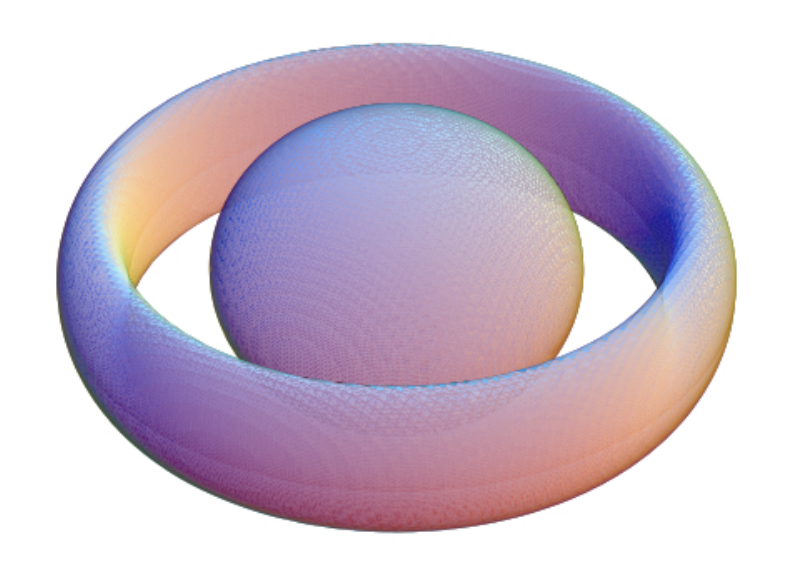} \qquad \qquad 
       \includegraphics[width=6.1cm]{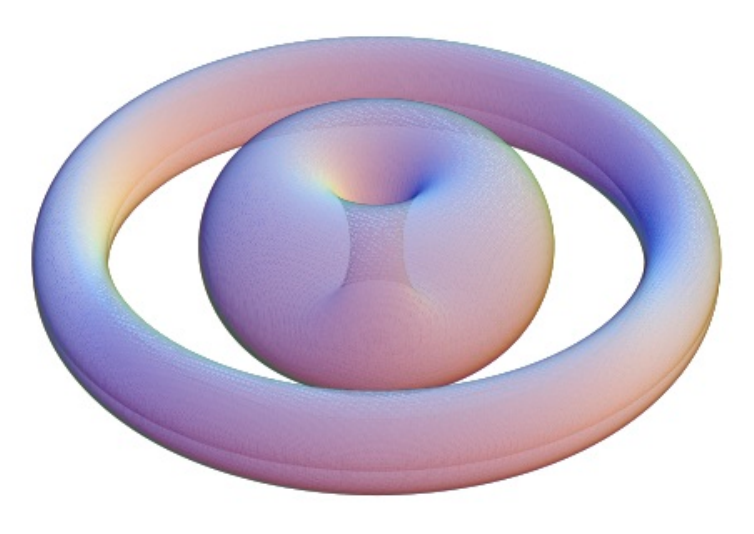}   
         \end{center}
 \caption{Left (right) panel: Saturn-like (di-ring-like) surfaces of constant energy density for the $m=1$ ($m=2$) Proca star exhibited in Fig.~\ref{PS1} (Fig.~\ref{PS2}). The corresponding energy density is $0.011$ ($0.008$). We emphasize these are not embedding diagrams; rather we defined Cartesian coordinates regarding the $r,\theta,\varphi$ coordinate system used here as standard spherical coordinates.}
  \label{3D}
\end{figure}

Although spinning Proca stars are quite similar to spinning scalar boson stars in many aspects, the energy and angular momentum density of the former exhibit novel features with respect to the latter. Spinning scalar boson stars for generic $m\geqslant 1$ are often described as an effective mass torus in general relativity~\cite{Schunck:1996he}, since surfaces of constant energy density present a toroidal topology sufficiently close to the centre of the star (see $e.g.$ the plots in~\cite{Herdeiro:2014ima}).  Spinning Proca stars, on the other hand, have a different structure for $m=1$ and $m>1$ as shown in Figs.~\ref{PS1}--\ref{3D} for illustrative cases (with $w=0.8$ and along the first branch for all examples). For $m=1$ the Proca star's energy density has a maximum at the origin and a second maximum (smaller) at some radial distance, thus presenting a composite-like structure, $cf.$ Fig.~\ref{PS1} (top left panel): instead of being toroidal some constant energy surfaces are \textit{Saturn-like} - Fig.~\ref{3D} (left panel). The angular momentum density, on the other hand, is zero at the origin and has two local positive maxima at some radii and one local negative minimum between them -- Fig.~\ref{PS1} (top right panel); in particular this means there is a counter-rotating toroidal-like region. For $m>1$ the Proca star's energy density vanishes at the origin and two local maxima arise at different radial values, $cf.$ Fig.~\ref{PS2} (top left panel). Thus some constant energy density surfaces are \textit{di-ring-like} - Fig.~\ref{3D} (right panel). The angular momentum density is similar to the $m=1$ case -- Fig.~\ref{PS2} (top right panel).

\begin{figure}[h!]
  \begin{center}
    \includegraphics[width=8.1cm]{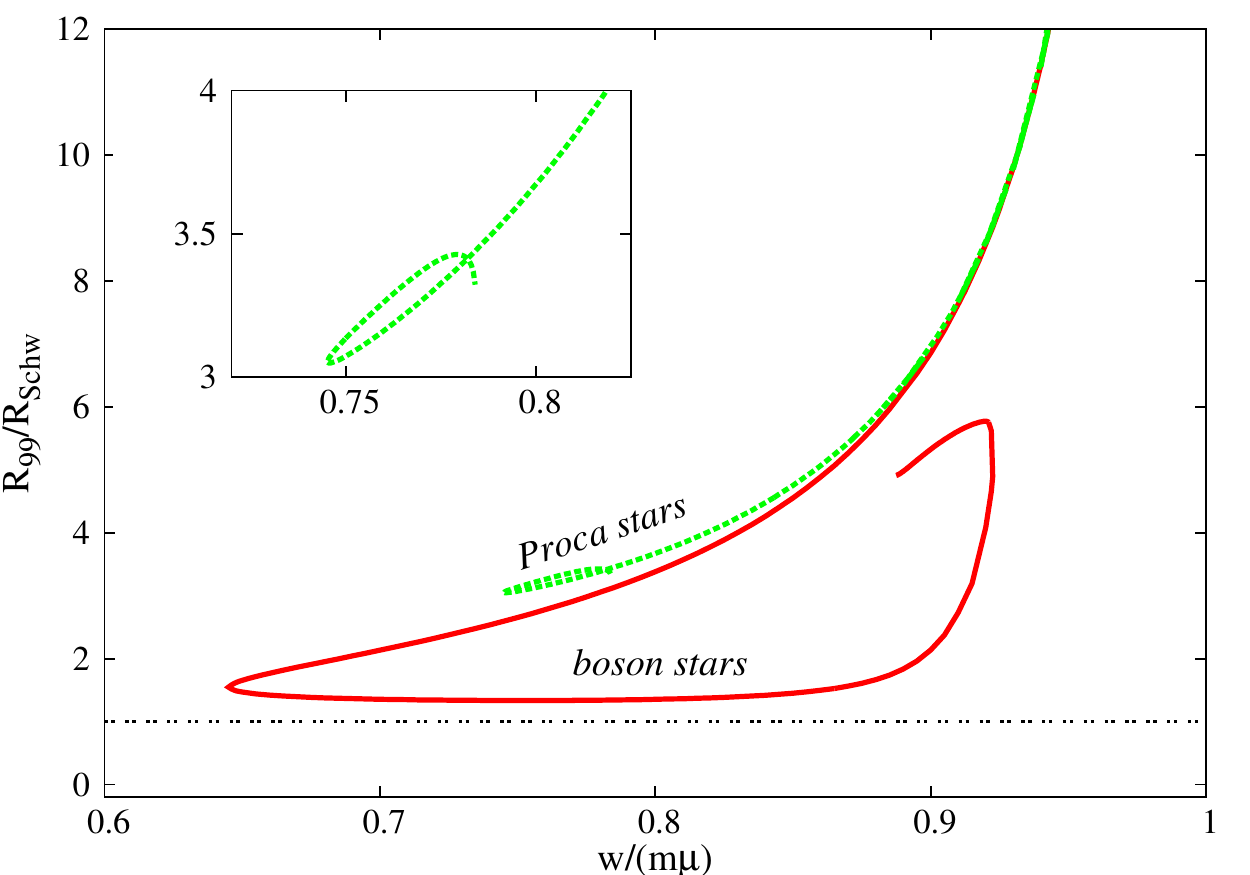}  
         \end{center}
 \caption{Inverse compactness of Proca stars compared to that of the scalar boson stars with $m=1$, defined in~\eqref{compactness}. The inset shows a detail of the Proca stars curve.}
  \label{compactnessfig}
\end{figure}

Finally, we discuss how `compact' these Proca stars are. Proca stars, like their scalar cousins, have no surface, $i.e.$ the Proca field decays exponentially towards infinity. Thus, there is no unique definition of the Proca star's `radius'. To obtain an estimate we follow the discussion in~\cite{AmaroSeoane:2010qx,Herdeiro:2015gia}. Using the `perimeteral' radius, $i.e.$, a radial coordinate $R$ such that a circumference along the equatorial plane has perimeter $\simeq 2\pi R$,  we compute $R_{99}$, the perimeteral radius containing 99\% of the Proca star mass, $M_{99}$. Then, we define the inverse compactness by comparing $R_{99}$ with the Schwarzschild radius associated to 99\% of the Proca star's mass, $R_{Schw}=2M_{99}$:
\begin{equation}
{\rm Compactness}^{-1}\equiv  \frac{R_{99}}{2M_{99}} \ .
\label{compactness}
\end{equation}
The result for the inverse compactness of Proca stars with $m=1$ is exhibited in Figure~\ref{compactnessfig}. With this measure, the inverse compactness is always greater than unity; $i.e.$, Proca stars are less compact than BHs, as one would expect, but they are also less compact than comparable scalar boson stars.

\section{Kerr BHs with Proca Hair} 
\label{sec_kbhsph}
We are now finally ready to tackle KBHsPH. The parallelism with the scalar case for both the stationary clouds and the solitonic limit is striking and one anticipates a high degree of similarity also at the level of the hairy BH solutions. 

The metric ansatz for constructing KBHsPH is the same as it was used for KBHsSH in~\cite{Herdeiro:2014goa}, and is precisely of the form~\eqref{kerrnc} with~\eqref{n}, where now 
all four (unspecified) functions $F_0,F_1,F_2,W$ depend on $(r,\theta)$ and, again, $r_H$ is a constant. If $F_2$ is finite, then $r={\rm constant}$ surfaces are timelike for $r>r_H$ and become null for $r=r_H$. Thus, $r=r_H$ is the location of the event horizon if the metric is regular therein.

For $r_H=0$, this ansatz reduces to the one discussed in the previous section for Proca stars, except for the replacement~\eqref{ww}. The line element form used for Proca stars is useful to tackle the behaviour at the origin, whereas the one used for BHs is useful to tackle the behaviour on a rotating horizon wherein $W$ reduces to the horizon angular velocity, $\Omega_H$. Indeed, following null geodesic generators ($ds^2=0$) on the horizon ($r=r_H$), assuming $F_2$ is finite therein, implies $d\varphi=W(r_H)dt$ and thus $W(r_H)=\Omega_H$, the angular velocity as measured by the observer at infinity.

The Proca field ansatz is the same as for the stationary Proca clouds (and Proca stars up to the replacement~\eqref{ww}),~\eqref{procaclouds}. This, again, introduces two parameters: $w>0$, $m\in \mathbb{Z}$. As for Proca stars we shall focus here on $m=1$, and take the sychronization condition~\eqref{synchronization} that we can rewrite in this context as (for general $m$)
\begin{equation}
\frac{w}{m}=W(r_H) =\Omega_H\ . 
\label{synchronization2}
\end{equation}
This condition was deduced in the context of a test field on the Kerr background and can be related to the threshold of superradiance. But it also has a different origin. In Appendix~\ref{appendixb}, we present the Einstein tensor and the Proca energy-momentum tensor associated to the ansatz discussed in this section. A careful inspection of the components of the energy-momentum tensor that have inverse powers of $N$,\footnote{A similar analysis can be made at the level of the components in an orthonormal frame, with similar conclusions.} and hence may diverge at the horizon, shows that, taking into account~\eqref{bccloudshorizon}, finiteness of the energy-momentum tensor components presented at $r=r_H$ \textit{requires}
\begin{equation}
\frac{w-mW(r_H)}{N(r_H)} 
\end{equation}
to be finite and hence it requires \eqref{synchronization2} (the same can be observed in the Einstein equations presented in~\cite{Herdeiro:2015gia}). It is interesting to remark that this finiteness condition~\eqref{synchronization2} is not necessarily related to superradiance, as the higher dimensional examples in~\cite{Brihaye:2014nba,Herdeiro:2015kha} illustrate.

The Einstein-Proca equations are solved with the following boundary conditions (which again we have found to be compatible with an approximate construction of the solutions
on the boundary of the domain of integration):  
\begin{description}
\item[i)] at infinity, the same as for Proca stars,~\eqref{bccloudslarge} and~\eqref{bcstarslarge};
\item[ii)] on the symmetry axis, the same as for Proca stars,~\eqref{bccloudsaxis} and~\eqref{bcstarsaxis};
 
\item[iii)] at the horizon, using again the new radial coordinate $x=\sqrt{r^2-r_H^2}$, a power series expansion near $x=0$ implies~\eqref{bccloudshorizon}, together with
\begin{equation}
\partial_x F_i\big|_{x=0}=0 \ , \qquad W\big|_{x=0}=\Omega_H \ .
\end{equation}
\end{description}

The Einstein-Proca equations for KBHsPH are quite involved (Appendix B). They are solved numerically, subject to the above boundary conditions, by  using the elliptic PDE solver \textsc{fidisol/cadsol}~\cite{schoen} 
based on a finite differences method in conjunction with the Newton-Raphson procedure. 
A description of the method for the case of KBHsSH can be found in~\cite{Herdeiro:2015gia}. 
The procedure in the case at hand is analogous.

\subsection{Physical Quantities}
\label{subsec_II}
In the following we shall describe some physical quantities that will be monitored from the numerical solutions we have obtained.  The ADM mass, $M$, and ADM angular momentum, $J$, are read off from the asymptotic expansion of the appropriate metric components:
\begin{equation}
\label{asym}
g_{tt} =-1+\frac{2M}{r}+\dots \ ,\qquad ~~g_{\varphi t}=-\frac{2J}{r}\sin^2\theta+\dots \ . \ \ \ 
\end{equation}
We also compute the horizon mass and angular momentum by using the appropriate Komar integrals associated to the corresponding Killing vector fields ${\bf k}$ and ${\bf m}$:
\begin{equation}
M_H=-\frac{1}{8\pi}\oint_{\mathcal{H}}dS_{\alpha\beta}D^\alpha k^\beta \ , \qquad 
J_H=\frac{1}{16\pi}\oint_{\mathcal{H}}dS_{\alpha\beta}D^\alpha m^\beta \ .
\end{equation}
Of course, $M$ and $J$ can also be computed as Komar integrals at infinity. Then, applying Gauss's law, one obtains a relation with $M_H$ and $J_H$ together with volume integrals on a spacelike surface with a boundary at the (spatial section of the) horizon. By making use of the Killing identity and the Einstein equations one obtains:
\begin{equation}
M=M_H-2\int_{\Sigma}dS_{\alpha}\left(T^\alpha_\beta k^\beta-\frac{1}{2}Tk^\alpha\right) \equiv M_H+M^{(\mathcal{P})}
\end{equation}
This defines the energy stored in the Proca field (outside the horizon):
\begin{equation}
M^{({\cal P})}\equiv - \int_{\Sigma} dr d\theta d\varphi(2T_t^t-T_\alpha^\alpha) \sqrt{-g} \ .
\label{ed}
\end{equation}
Proceeding similarly for the angular momentum one obtains:
\begin{equation}
J=J_H+J^{(\mathcal{P})} \ , \ \qquad  J^{({\cal P})}\equiv  \int_{\Sigma} dr d\theta d\varphi T^t_\varphi \sqrt{-g} \ ,
\label{amd}
\end{equation}
which defines the angular momentum stored in the Proca field. At this point, an interesting distinction arises, with respect to the scalar case. 
 Whereas for KBHsSH the angular momentum stored in the scalar field relates to the Noether charge in precisely the same way as for rotating scalar boson stars $J^{(\Psi)}=mQ$, for KBHsPH the relation between $J^{(\mathcal{P})} $ and  the Noether charge~\eqref{q} includes an extra boundary term (see Appendix~\ref{appendixc} and eq.~\eqref{JQBHs})
\begin{equation}
\label{nr1}
J^{(\mathcal{P})}=mQ+ \oint_\mathcal{H}  ({\mathcal{A}}_\varphi \bar{ {\mathcal{F}}}^{r t}+\bar{\mathcal{A}}_\varphi { {\mathcal{F}}}^{r t}  ) dS_r \ ,
\end{equation}
which generalizes relation~\eqref{amnc} to the case of hairy BHs. 
A similar relation can be written for $M^{({\cal P})}$ (see Appendix \ref{appendixc}
and eq. (\ref{sup1}))
\begin{equation}
\label{nr2}
M^{({\cal P})}=2w Q
-\mu^2  {\cal U}
+ \oint_\mathcal{H}  
\left[
\frac{1}{2} 
\left(
{\mathcal{A}}_\beta \bar{ {\mathcal{F}}}^{r \beta}+\bar{\mathcal{A}}_\beta { {\mathcal{F}}}^{r \beta}
\right)
-\left({\mathcal{A}}_t \bar{ {\mathcal{F}}}^{r t}+\bar{\mathcal{A}}_t { {\mathcal{F}}}^{r t}
\right)  
\right] dS_r \ ,
\end{equation}
with
\begin{equation}
\label{U}
{\cal U}\equiv \int _\Sigma  dr d\theta d\varphi  {\mathcal{A}}_\alpha \bar {\mathcal{A}}^\alpha \sqrt{-g}\ .
\end{equation}

The horizon temperature and event horizon area of the KBHsPH solutions are computed by standard relations, that specialize to: 
\begin{eqnarray}
\label{THAH}
T_H=\frac{1}{4\pi r_H}e^{(F_0-F_1)|_{r=r_H}} \ , \qquad
 A_H=2\pi r_H^2 \int_0^\pi d\theta \sin \theta  e^{(F_1+F_2)|_{r=r_H}}\ . 
 \end{eqnarray}
Then, the ADM quantities $M,J$ are related with $T_H,S,Q,M^{({\cal P})}$, where $S=A_H/4$ is the horizon entropy, through a Smarr formula  
\begin{eqnarray}
\label{smarr} 
M=2 T_H S +2\Omega_H J_H+ M^{({\cal P})} \ .
\end{eqnarray}
Also, the variation of $M$ can be expressed by the first law:
\begin{equation}
\label{fl}
dM=T_H dS +\Omega_H dJ \ .
\end{equation}
We note that
by making use of the relations
(\ref{nr1})
and
(\ref{nr2}),
 the Smarr formula (\ref{smarr})
can be written in a Kerr-like form
\begin{eqnarray}
\label{smarr-new1} 
M=2 T_H S +2\Omega_H J-\mu^2 {\cal U} \ ,
\end{eqnarray}
which renders explicit the fact that the solutions are supported by
a nonzero mass term of the Proca field.

Finally, we observe that Proca stars
satisfy a simple relation, which results again from 
(\ref{nr1}),
(\ref{nr2}):\footnote{One can similarly show that KBHsSH and scalar boson stars satisfy relations analogous to
(\ref{smarr-new1}) and  (\ref{smarr-new2}), respectively.} 
\begin{eqnarray}
\label{smarr-new2} 
M=2 w Q-\mu^2 {\cal U}=2\frac{w}{m} J-\mu^2 {\cal U}\ .
\end{eqnarray}

\subsection{The domain of existence and phase space}
\label{subsec_III}
\begin{figure}[h!]
  \begin{center}
    \includegraphics[width=8.1cm]{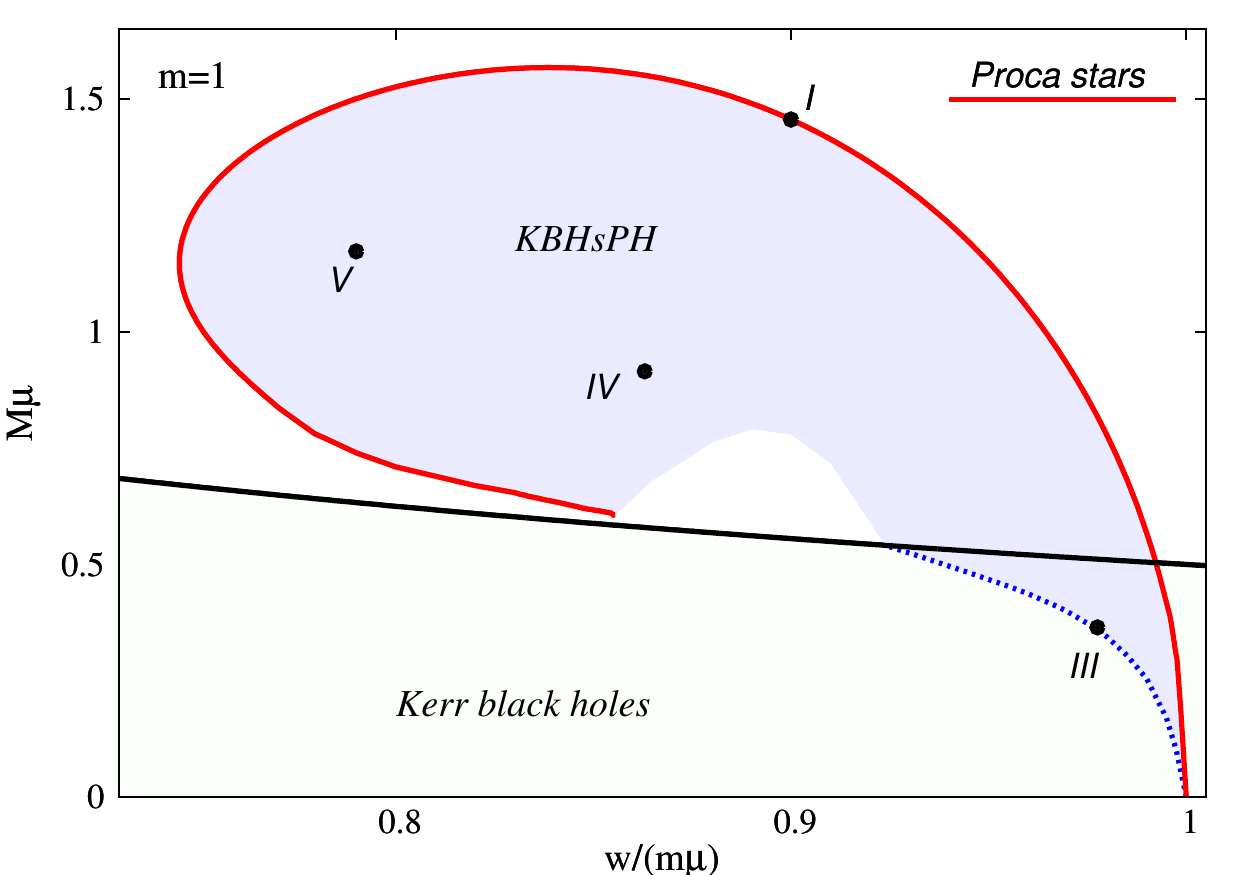}
      \includegraphics[width=8.1cm]{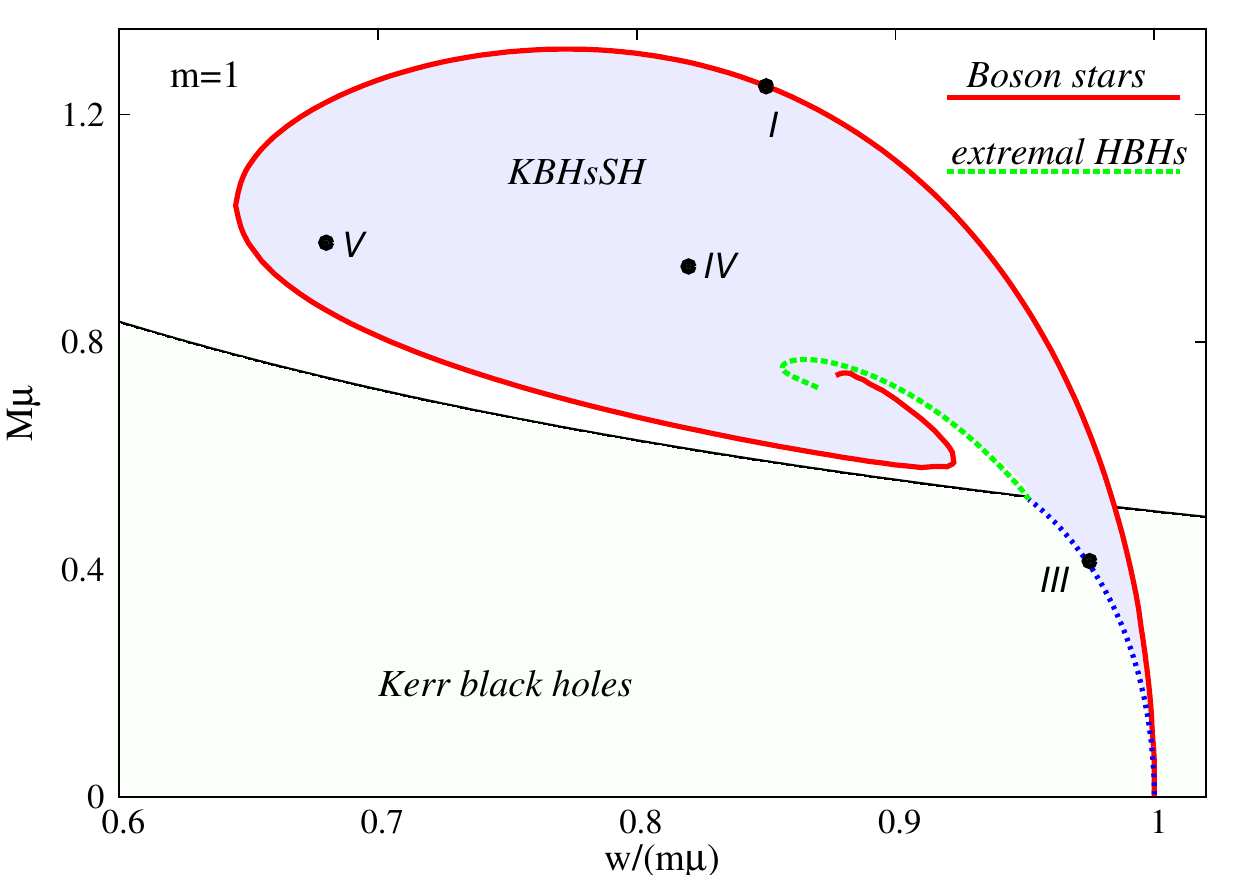}
  \end{center}
  \caption{ADM mass $vs.$ frequency $w$ diagram for $m=1$ KBHsPH (left panel) and KBHsSH (right panel). The red solid lines correspond to the solitonic limit (Proca stars and scalar boson stars, respectively, already shown in Fig.~\ref{clouds}). The blue dotted lines are the Kerr limit, also shown in Fig.~\ref{clouds}. Kerr solutions exist below the black solid line, which corresponds to extremal Kerr solutions. The hairy BHs exist in the blue shaded region. Points I,III,IV,V, in each case, correspond to specific solutions for which the numerical data is publicly available~\cite{datakbhph,datakbhsh}. The right panel also shows the extremal hairy BHs (green dashed) line.}
  \label{figdomain}
\end{figure}

We have scanned the domain of existence of KBHsPH by varying $r_H$ for fixed $w$ lines (or vice-versa), 
in between the minimum frequency 
$w_{\rm min}/\mu=0.7453$ and the maximal one $w=\mu$. 
The result for the $m=1$ family of KBHsPH is shown in Fig.~\ref{figdomain} (left panel), 
together with the analogous family of KBHsSH (right panel), the former obtained from over five thousand numerical points.

Based on the discussions of KBHsSH~\cite{Herdeiro:2014goa,Herdeiro:2015gia,Herdeiro:2015tia}, and as already partly discussed, the domain of existence of KBHsPH should be bounded by three lines: the Proca clouds existence line discussed in Section~\ref{sec_clouds}, the Proca star line discussed in Section~\ref{sec_stars} and the line of extremal KBHsPH ($i.e.$ zero temperature). So far, the last of the three were only obtained by extrapolating to $T_H=0$ the non-extremal solutions, as our attempts to construct the extremal KBHsPH solutions by directly solving the Einstein-Proca field equations were unsuccessful (unlike the scalar case, as reported in~\cite{Herdeiro:2015gia}). For this reason we have chosen not to display this line in Fig.~\ref{figdomain}, for the Proca case.  Another technical difficulty arises in trying to connect the set of (extrapolated) extremal solutions with the set of Proca stars. As for the case of KBHsSH, these two curves are likely to meet in a critical point at the center of the 
Proca stars spiral; however, validation of this hypothesis is a numerical challenge (also for KBHsSH).

Concerning numerical errors, the PDE solver we have used provides error estimates for each unknown function, which allows judging the quality of the computed solution. The numerical error for the solutions reported in
this work is estimated to be typically $<10^{-3}$. As a further check of the numerical procedure, we have verified that the families of solutions  satisfy with a very good accuracy the first law of thermodynamics and also the Smarr relation, typically at that same order. We have also monitored the violation of the gauge condition together with the constraint Einstein equations; typically, these provide much lower estimates for the numerical errors. As a comparative comment, the overall quality of the solutions is, however,  not as high for KBHsPH as for KBHsSH.
Additionally, the source of the difficulties we have encountered in constructing extremal and close to extremal solutions are absent in the scalar case. Typically, for the Proca case, the solver stops to converge in the near extremal case,
 although the error estimates for the last solutions is still small. It is likely that another metric parametrization is required to tackle this issue. We also remark that  there may be a more involved landscape of excited solutions in view of the four vector potentials.\footnote{ 
In fact, we have observed that the solver frequently ``jumps"  to one of these excited configurations 
which is not too far in the parameter space.}

In Fig.~\ref{figdomain} we have singled out four particular solutions for each case, denoted I,III,IV and V. The numerical data for these four solutions, together with the data for a vacuum Kerr solution with the same ADM mass and angular momentum as that of configuration III, for each case, has been made publicly available for community use~\cite{datakbhph,datakbhsh}. The corresponding parameters are detailed in Appendix~\ref{appendixd}.

In Fig.~\ref{fig2} we exhibit the phase space, $i.e.$ ADM mass $vs.$ ADM angular momentum diagram for $m=1$ solutions of KBHsPH (left panel) and as a comparison, the corresponding diagram for KBHsSH (right panel). The two plots are quite similar and the features we wish to emphasize is that, as for the scalar case, one observes violation of the Kerr bound (in terms of ADM quantities) and non-uniqueness, $i.e$ there are both hairy and vacuum Kerr BHs with the same ADM mass and angular momentum ($cf.$ Appendix~\ref{appendixd}).

%
\begin{figure}[h!]
  \begin{center}
    \includegraphics[width=8.1cm]{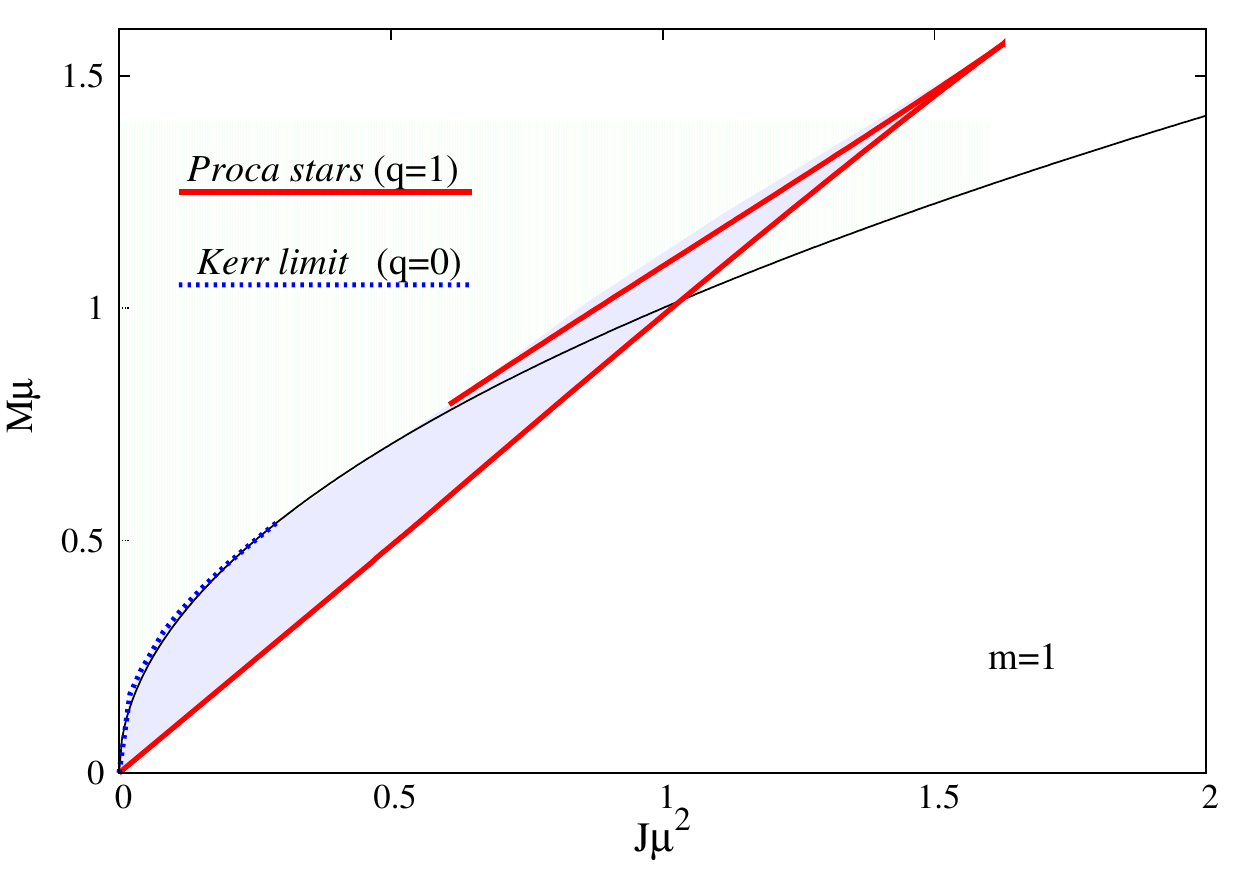}
      \includegraphics[width=8.1cm]{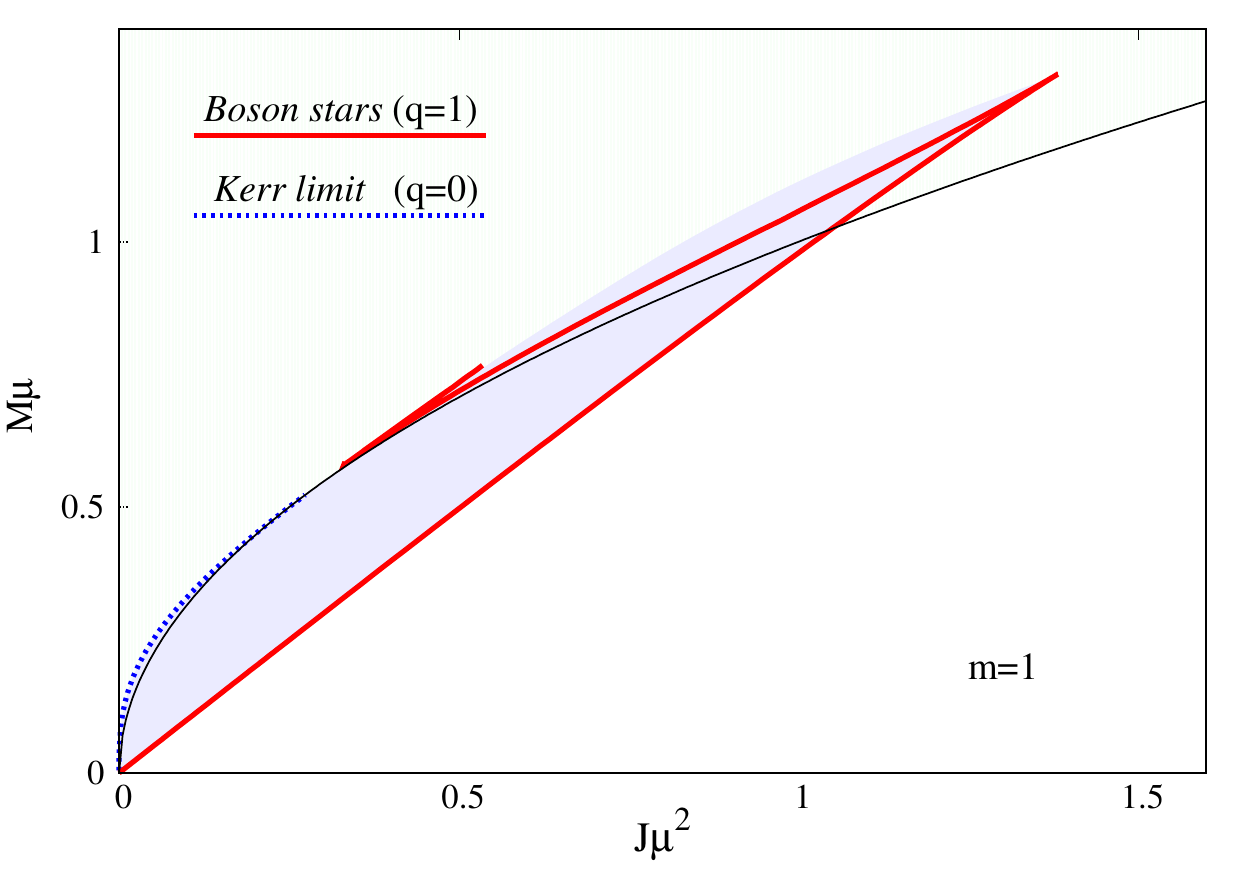}
  \end{center}
  \caption{ADM mass $vs.$ ADM angular momentum diagram for $m=1$ KBHsPH (left panel) and KBHsSH (right panel), in units of the field mass. The black solid line corresponds to extremal Kerr solutions; non extremal BHs exist above this line. The red solid line is for Proca (scalar boson) stars in the left (right) panel. The blued dotted line is the existence line, denoting Kerr BHs that support Proca (scalar) clouds. The blue shaded region is the domain of existence of KBHsPH (KBHsSH).}
  \label{fig2}
\end{figure}

\begin{figure}[h!]
  \begin{center}
    \includegraphics[width=8.1cm]{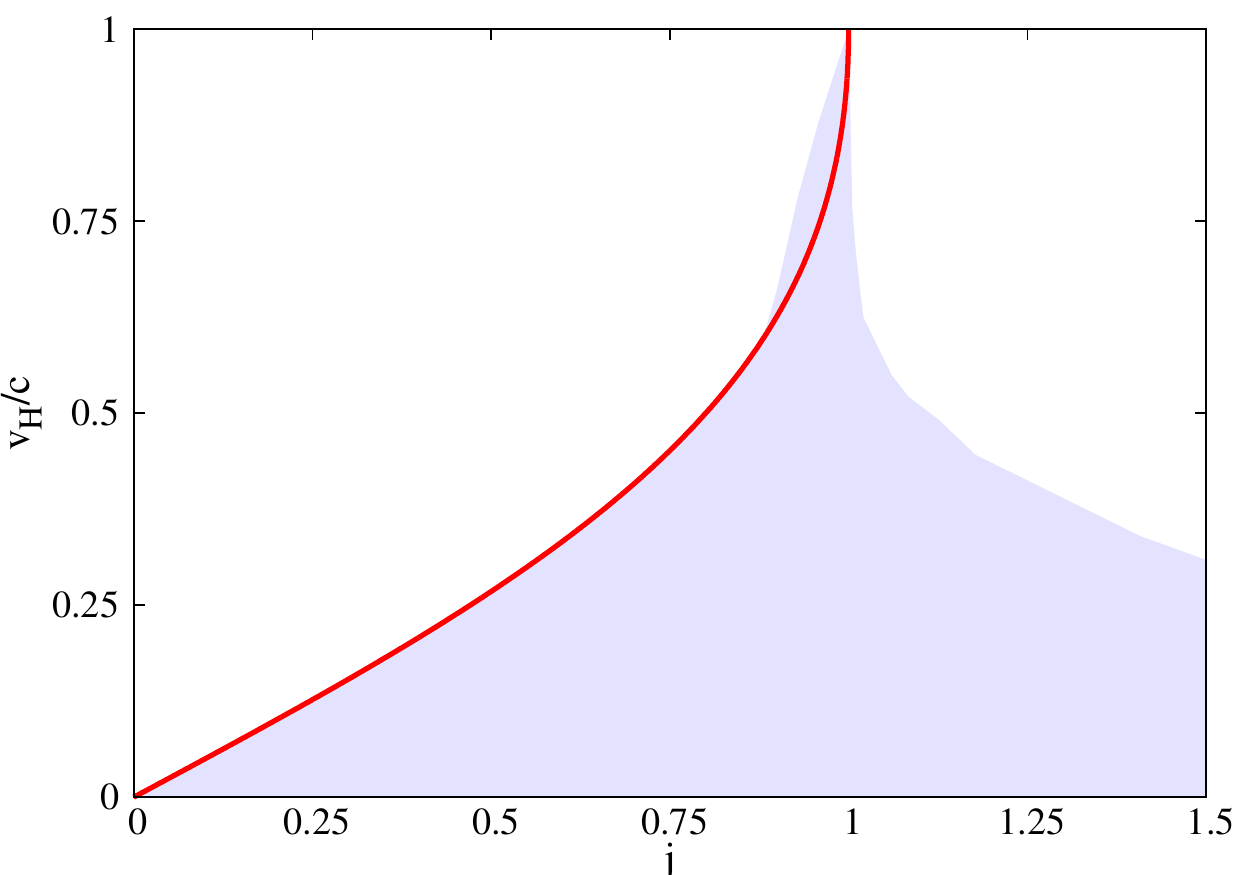}
      \includegraphics[width=8.1cm]{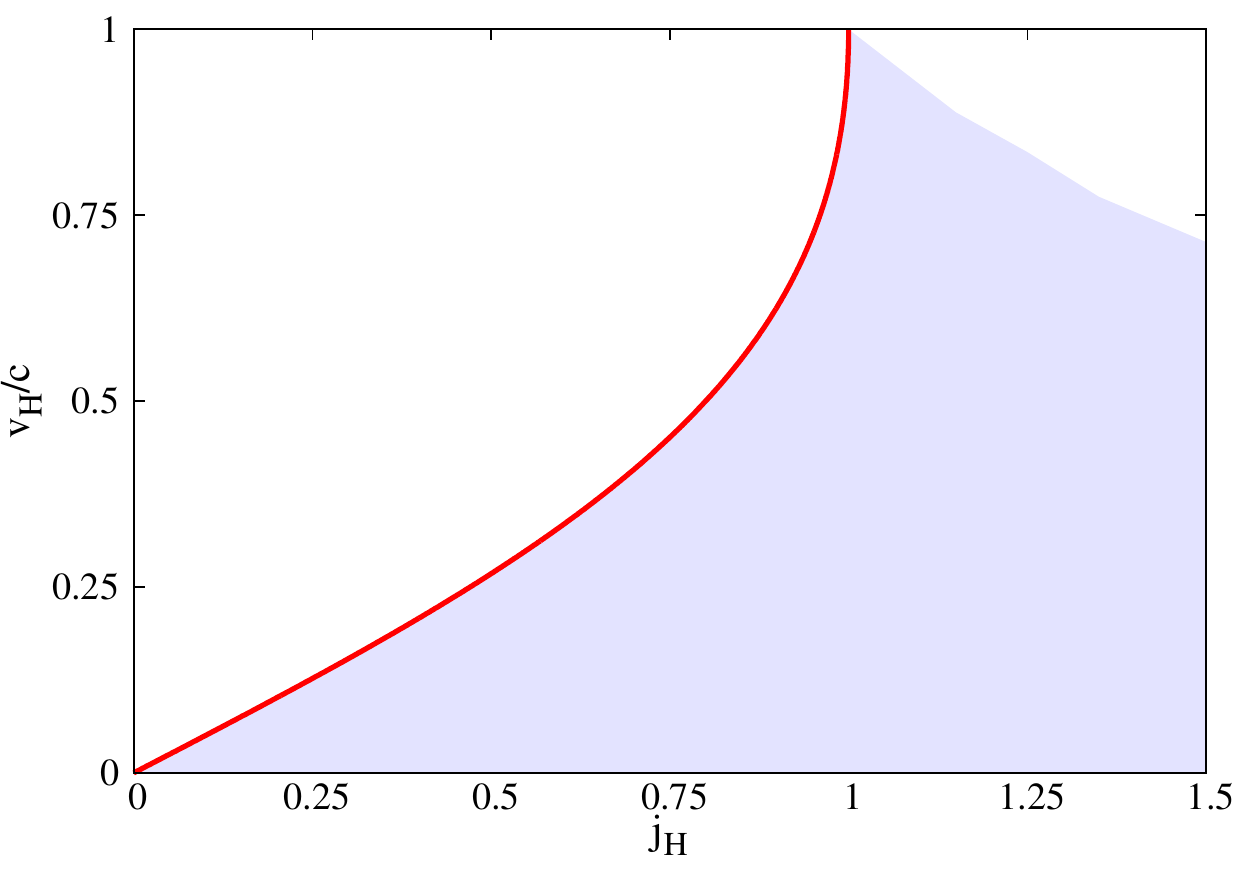}
  \end{center}
 \caption{Linear velocity of the horizon normalized to the speed of light, $v_H$, versus:  (left panel)  the ADM dimensionless spin parameter $j\equiv Jc/GM^2$, where $M,J$ are the ADM mass and angular momentum; (right panel) the horizon dimensionless spin parameter $j_H\equiv J_Hc/GM_H^2$, where $M_H,J_H$ are the horizon mass and angular momentum. Here we have reinstated $c,G$. The red solid line corresponds to vacuum Kerr and the shaded area is filled by KBHsPH.}
  \label{vconjecture}
\end{figure}

The violation of the Kerr bound also occurs in terms of \textit{horizon} quantities, as shown in Fig.~\ref{vconjecture} (right panel). For these solutions the conjecture put forward in~\cite{Herdeiro:2015moa} concerning the horizon linear velocity $v_H$, as defined therein, holds: despite violating the Kerr bound both in terms of ADM and horizon quantities, $v_H$ never exceeds the speed of light.  We recall $v_H$ is defined as follows, for asymptotically flat, stationary and axi-symmetric spacetimes. On a spatial section of the event horizon one computes the proper length of all closed orbits of ${\bf m}$. Let $L_{\rm max}$ be the maximum of all such proper lengths; the corresponding circumferencial radius, $R_c$, is $R_c\equiv {L_{\rm max}}/({2\pi})$. 
The horizon linear velocity is $v_H \equiv R_c \Omega_H$ \cite{Herdeiro:2015moa}.

\subsection{Energy distribution and horizon quantities}
\label{subsec_IV}
\begin{figure}[h!]
  \begin{center}
    \includegraphics[width=8.1cm]{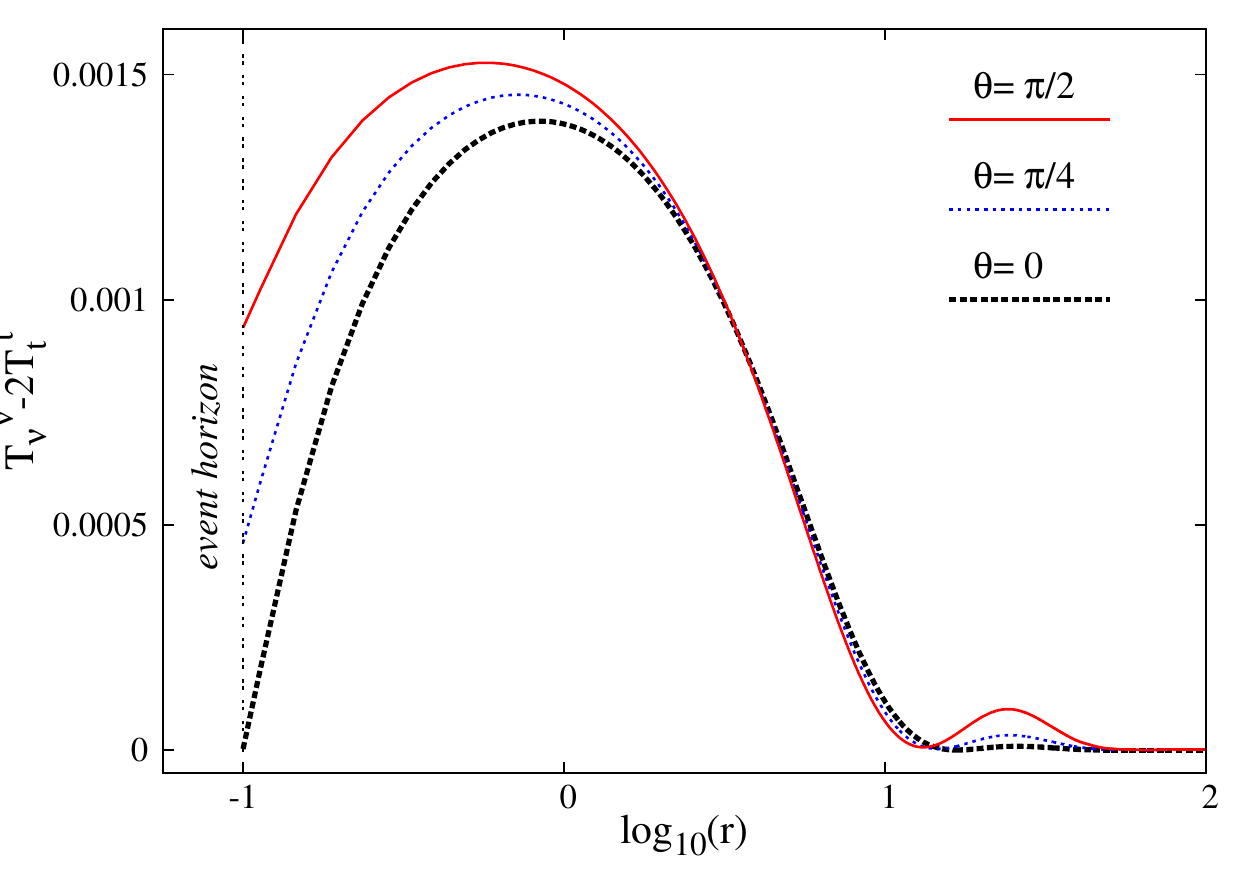}
      \includegraphics[width=8.1cm]{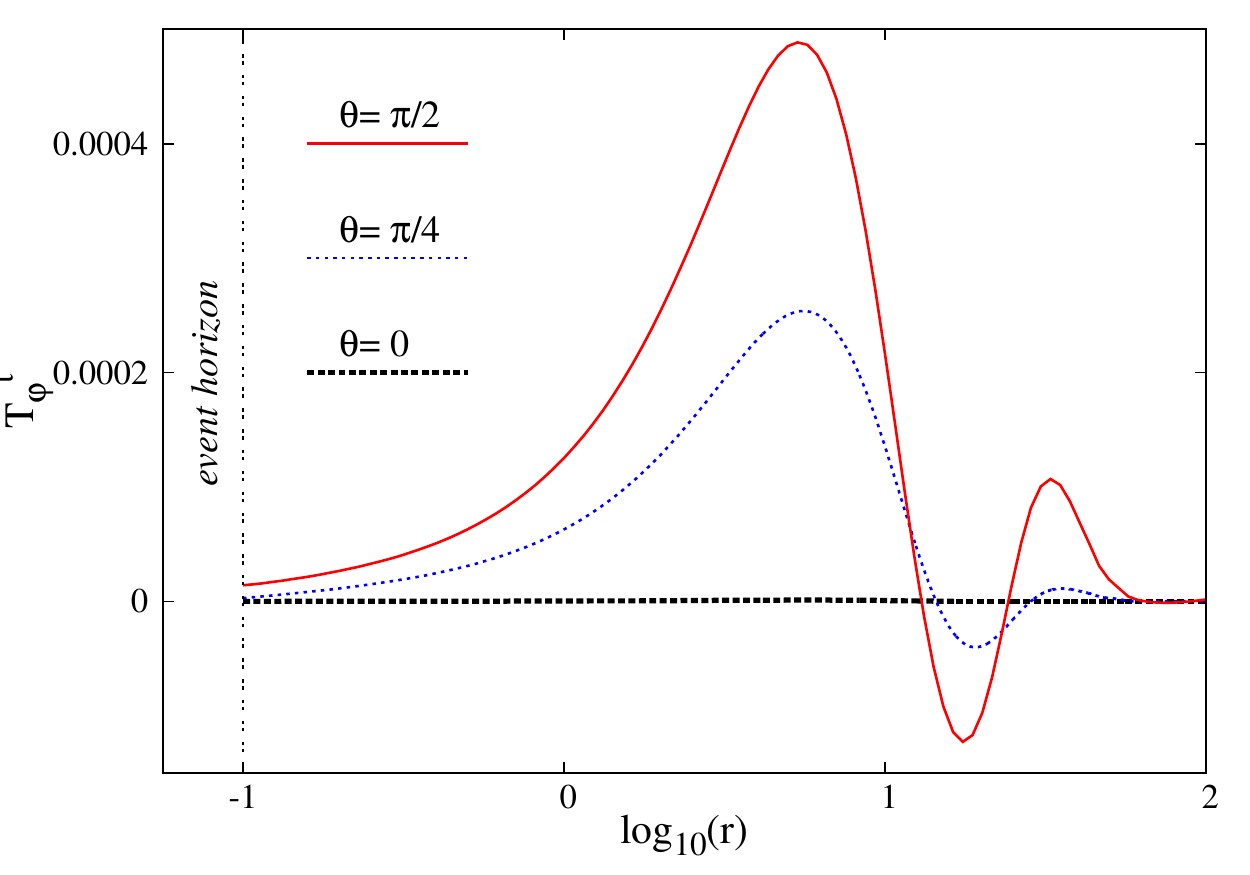}
  \end{center}
  \caption{Radial variation of the energy density, $cf.$~\eqref{ed} (left panel), and angular momentum density, $cf.$~\eqref{amd}  (right panel),  of the Proca field, for different constant $\theta$ sections of a KBHPH with $m=1$, $w=0.98\mu$, $r_H=0.1$,  $\mu M=0.701$ and $\mu^2J=0.652$.}
  \label{figenergybhs}
\end{figure}

\begin{figure}[h!]
  \begin{center}
    \includegraphics[width=6.1cm]{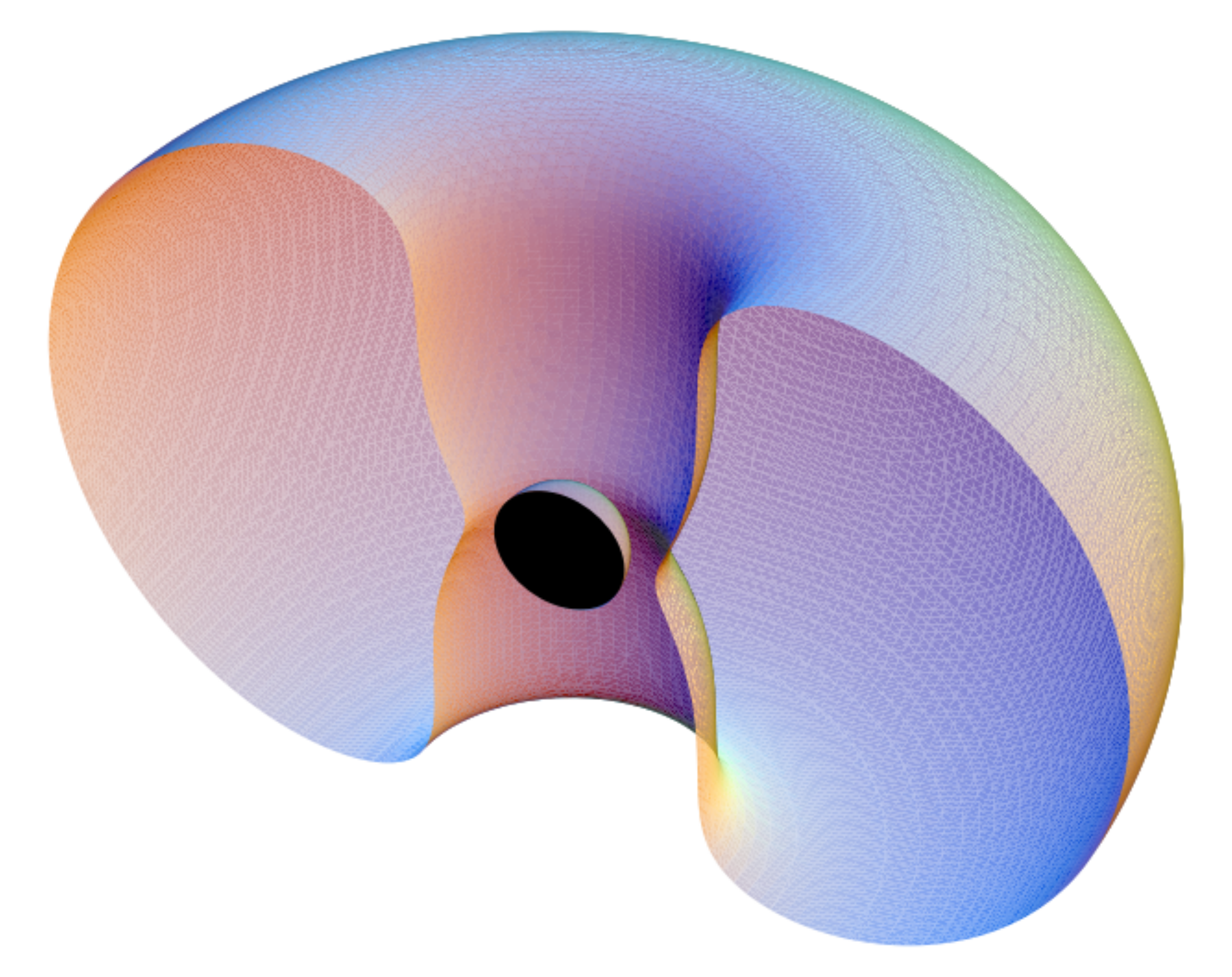}
  \end{center}
  \caption{One toroidal-like surface of constant energy density (corresponding to $0.00142$) for the same KBHPH displayed in Fig.~\ref{figenergybhs}. We also plot the spatial section of the event horizon in these coordinates (half-sphere with the black cross section).}
  \label{fig3Dbh}
\end{figure}

\begin{figure}[h!]
  \begin{center}
    \includegraphics[width=5.5cm]{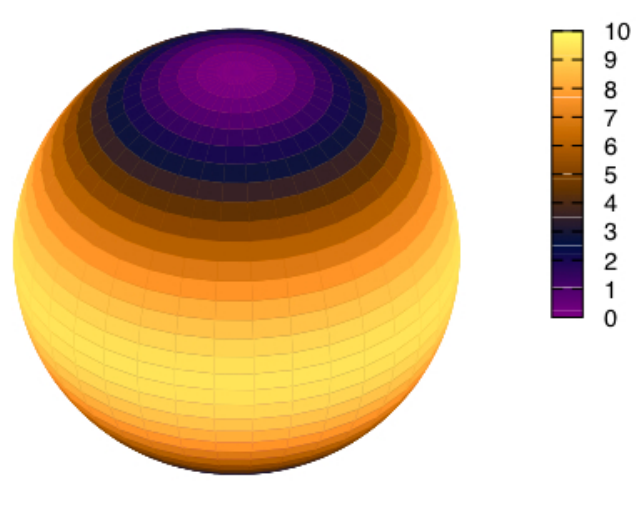}\qquad \qquad 
      \includegraphics[width=6.0cm]{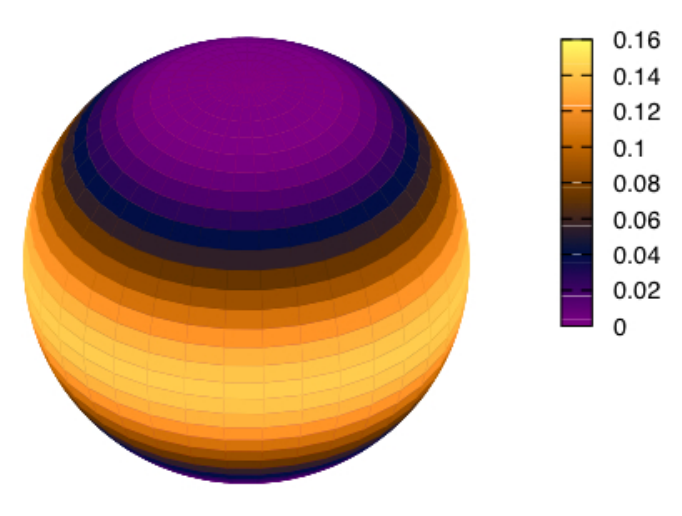}
  \end{center}
  \caption{Energy density, $cf.$~\eqref{ed} (left panel), and angular momentum density, $cf.$~\eqref{amd}  (right panel),  of the Proca field on the horizon for the same example of a KBHPH displayed in Fig.~\ref{figenergybhs}. The corresponding values were multiplied by $10^4$ for better visualization.}
  \label{horizoned}
\end{figure}

As for their scalar cousins, KBHsPH can be thought of as a bound state of a horizon with a Proca star. Thus, the matter energy density distribution around the horizon will resemble that of (some) Proca stars. In~Fig.~\ref{figenergybhs}  we exhibit the energy density and the angular momentum density as a function of the radial coordinate for different angular sections for an example of KBHPH. As for the Proca stars, both the energy density and the angular momentum density can have more than one maximum outside the horizon and the latter can also have regions with a different sign. Thus, outside KBHsPH there are counter-rotating regions. In Fig.~\ref{fig3Dbh} a constant Proca energy density surface is exhbited in a 3D plot. The behaviour of the energy density and angular momentum density on the horizon is more clearly seen in Fig.~\ref{horizoned}.

\begin{figure}[h!]
  \begin{center}
    \includegraphics[width=8.1cm]{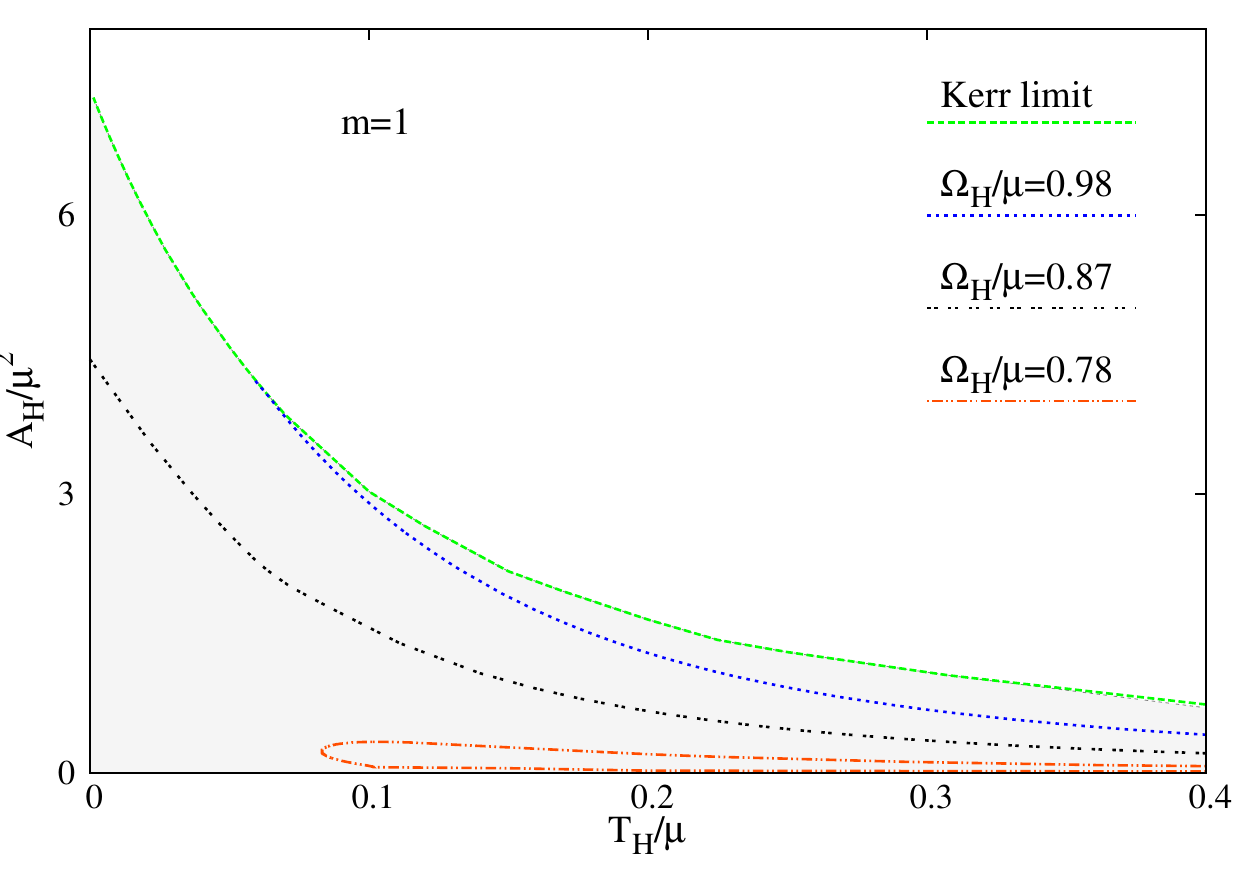}
      \includegraphics[width=8.1cm]{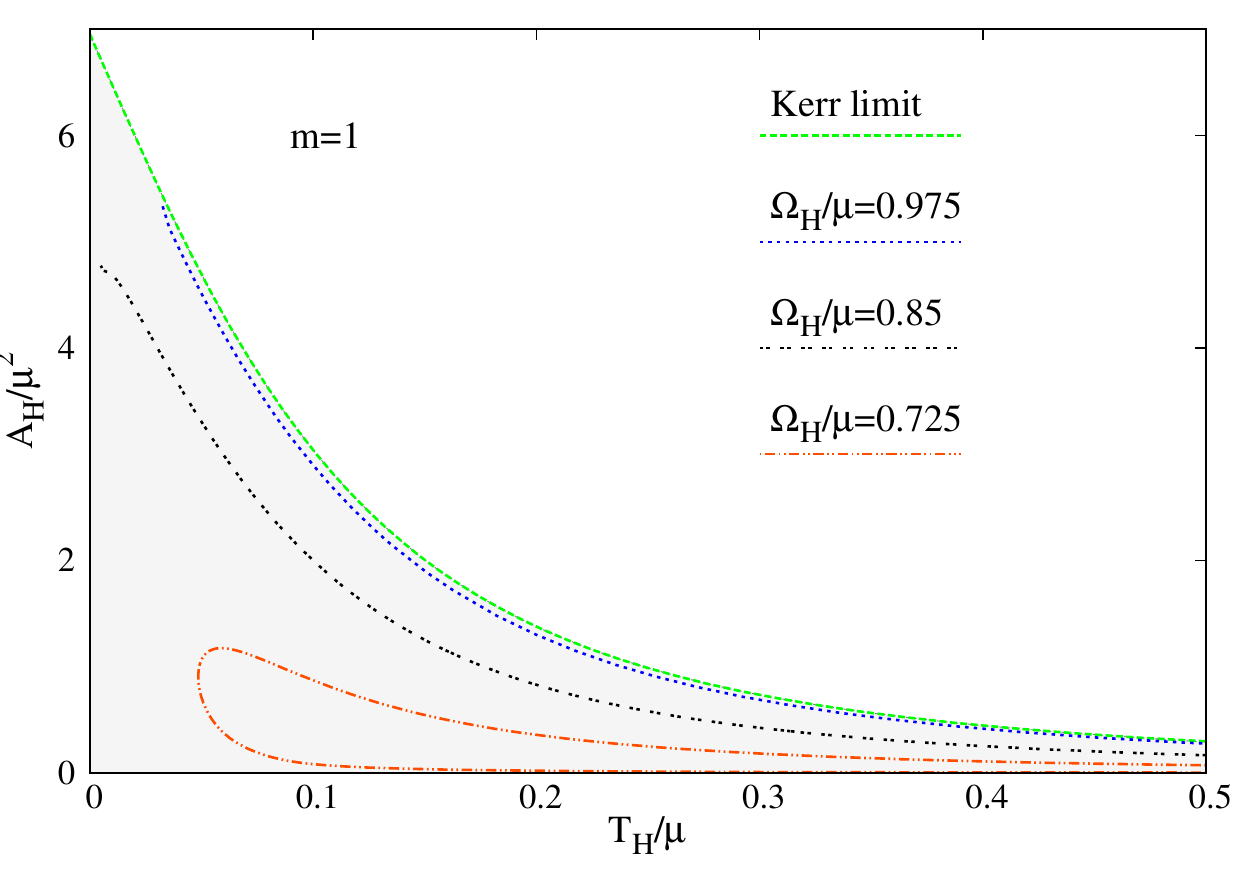}
  \end{center}
 \caption{Event horizon area $vs.$ temperature for KBHsPH (left panel) and KBHsSH (right panel), in units of $\mu$, for different constant angular velocity sets of solutions.}
  \label{temperature}
\end{figure}

Finally, in Fig.~\ref{temperature} we exhibit the variation of the horizon area with the horizon temperature along sequences of solutions with constant horizon angular velocity (or frequency). For both KBHsPH (left panel) and KBHsSH (right panel) one can see three different types of behaviour, which are easy to interpret referring back to Fig.~\ref{figdomain}. For large values of $\Omega_H$, the solutions interpolate between the Kerr existence line and the corresponding (Proca or scalar boson) star line (for which $T_H\rightarrow \infty$). For intermediate values of $\Omega_H$, the solutions interpolate between the extremal BHs line (for which $T_H\rightarrow 0$) and the corresponding star line. Finally, for sufficiently small values of $\Omega_H$, the solutions interpolate between two stars, and thus start and end for $T_H\rightarrow \infty$.

\section{Discussion}
\label{sec_discussion}
It has long been established that stationary, asymptotically flat BHs in Einstein's gravity minimally coupled to one or many real, Abelian Proca fields cannot have Proca hair. The basic theorem supporting this idea, due to Bekenstein~\cite{Bekenstein:1971hc,Bekenstein:1972ky}, assumes, however, that the Proca field inherits the spacetime isometries. In this paper we have shown that dropping this assumption Kerr BHs with Proca hair exist under two conditions:
\begin{description}
\item[i)] The Proca field is complex, or equivalently there are two real Proca fields with the same mass. Solutions in this paper can be, moreover, generalized to an arbitrary number of complex Proca fields (any even number of real Proca fields), without mutual interactions, and all of them minimally coupled to gravity. Here, however, we focus on a model with a single complex Proca field. 
\item[ii)] The complex Proca field has a harmonic time dependence, as in the ansatz~\eqref{procaclouds}, with the frequency and azimuthal harmonic index obeying the synchronization condition~\eqref{synchronization}.
\end{description}
These two assumptions, together, allow the two real Proca fields to oscillate, with the same frequency but opposite phases, hence cancelling out gravitational radiation emission (as well as Proca radiation emission). It remains as an open question if the same could be achieved with a single real Proca field, especially in view of the result in~\cite{Wang:2015fgp}, since such real Proca field already has two independent modes. 

\bigskip

The existence of  KBHsPH -- to the best of our knowledge the first example of (fully non-linear)   
BHs with (Abelian) vector hair -- is anchored  in the synchronization/superradiance zero mode condition
($i.e.$ the field should co-rotate with the  black hole horizon). 
All previously constructed examples which employed this mechanism have scalar hair, 
both in four spacetime dimensions~\cite{Herdeiro:2014goa,Herdeiro:2015gia,Kleihaus:2015iea,Herdeiro:2015tia} and in higher dimensions~\cite{Brihaye:2014nba,Herdeiro:2015kha}, 
including the example in five dimensional asymptotically Anti-de-Sitter space found in~\cite{Dias:2011at}. 
This further shows the generality of the mechanism and lends support to the conjecture in~\cite{Herdeiro:2014goa,Herdeiro:2014ima}.

We also remark that  the Proca model considered here can be regarded as a proxy 
for more realistic models with a gauged scalar field, where the gauge fields acquire a mass \textit{dynamically}, via the Higgs mechanism. A familiar example in this direction is the non-Abelian Proca model,
whose solutions contain already all basic properties of the Yang-Mills--Higgs sphalerons
in the Standard Model \cite{Greene:1992fw}. 
As such, the results in this work suggest that one should
reconsider the no-hair theorem for the Abelian-Higgs model~\cite{Adler:1978dp}.

\bigskip

Several direct generalizations/applications of these solutions are possible. 
At the level of constructing further 
solutions, we anticipate that 
$(i)$ self-interacting Proca hair will lead to new solutions, 
which, if the scalar field case is a good guide~\cite{Herdeiro:2015tia}, can have a much larger ADM mass (but not horizon mass)
and
$(ii)$ hybrid solutions with scalar plus Proca hair are possible. 
At the level of possible astrophysics phenomenology, 
it would be interesting to look in detail to the geodesic flow, 
in particular to the frequency at the innermost stable circular orbit (ISCO), 
quadrupoles as well as to the lensing and shadows of these new BHs,
following~\cite{Cunha:2015yba} (see also the review~\cite{Johannsen:2015mdd}). 
Work in this direction is underway.

\bigskip

 Finally, it is still a common place to find in the current literature statements that stationary BHs in GR are described solely by mass, angular momentum and charge. We want to emphasize that the examples of Kerr BHs with scalar and Proca hair show that this is \textit{not true as a generic statement for GR, even if physical matter -- i.e. obeying all energy conditions -- is required.} These examples show that Noether charges, rather than charges associated to Gauss laws, are also permitted in non-pathological stationary, asymptotically flat, BH solutions. The main outstanding questions is if in a real dynamical process these Noether charges can survive.

\vspace{0.5cm} 
\noindent
\section*{Acknowledgements}
We would like to thank Richard Brito and Vitor Cardoso for a fruitful collaboration on Proca stars. We also thank J. Rosa, M. Sampaio and M. Wang for discussions on Proca fields. C. H. and E. R. acknowledge funding from the FCT-IF programme. H.R. is supported by the grant PD/BD/109532/2015 under the MAP-Fis Ph.D. programme. This work was partially supported by  the  H2020-MSCA-RISE-2015 Grant No.  StronGrHEP-690904, and by the CIDMA project UID/MAT/04106/2013. Computations were performed at the Blafis cluster, in Aveiro University.
 
 \bigskip
 
\appendix

\section{Spheroidal prolate coordinates for Kerr}
\label{prolatecoordinates}
The new coordinate system for Kerr~\eqref{kerrnc}, with the functions~\eqref{functionsKerr}, first introduced in~\cite{Herdeiro:2015gia}, actually reduce to spheroidal \textit{prolate} coordinates in the Minkowski space limit, but with a non-standard radial coordinate. To see this, we observe that, from~\eqref{Kerr2}, $M=0$ occurs when $r_H=-2b$. Then, from the expressions~\eqref{functionsKerr}, the metric~\eqref{kerrnc} becomes
\begin{eqnarray}
ds^2=-dt^2+\left[N(r)+\frac{b^2}{r^2}\sin^2\theta\right]\left[\frac{dr^2}{N(r)}+r^2d\theta^2\right]
+N(r)r^2\sin^2\theta d\varphi^2 \ , \qquad N(r)\equiv 1+\frac{2b}{r} \ .
\end{eqnarray}
This can be converted to the standard Minkowski Cartesian quadratic form $ds^2=-dt^2+dx^2+dy^2+dz^2$ by the spatial coordinate transformation
\begin{equation}
\left\{
\begin{array}{l}
x=r\sqrt{N(r)}\sin\theta\cos\varphi \ , \\
y=r\sqrt{N(r)}\sin\theta\sin\varphi \ , \\
z=(r+b)\cos\theta \ .
\end{array}
\right.
\end{equation}
A surface with $r=$constant is, in Cartesian coordinates,
\begin{equation}
\frac{x^2+y^2}{\bar{r}^2}+\frac{z^2}{\bar{r}^2+b^2}=1 \ ,
\end{equation}
where $\bar{r}=r\sqrt{N(r)}$. This is a prolate spheroid. It is interesting that KBHsSH and KBHsPH seem to prefer prolate spheroidal coordinates rather than the oblate spheroidal coordinates so well adapted to Kerr (in the Boyer-Lindquist form).

\section{Einstein-Proca equations of motion}
\label{appendixb}

Here we provide explicit expressions for the equations of motion solved by KBHsPH. The components of the Einstein tensor for the ansatz~\eqref{kerrnc} are:

\begin{eqnarray}
&&  4r^2  e^{2F_1}{\bf G_{tt}}= 4 
N \left[r^2 ({F_1}_{,rr}+ {F_2}_{,r}^2+
{F_2}_{,rr})
+r( 
{F_1}_{,r}+3  {F_2}_{,r})+1\right] + 2 r N'  \left(r 
{F_1}_{,r}+r {F_2}_{,r}+2\right)+4 {F_1}_{,\theta\theta}\nonumber \\
&& +4  
{F_2}_{,\theta}^2+4  {F_2}_{,\theta\theta}
+8
\cot \theta {F_2}_{,\theta}-4 
 +r^4 \sin ^2\theta e^{2 ({F_2}-F_0)} 
 \left\{
 \frac{2W W_{,r}}{r}(- r {F_0}_{,r} +3r{F_2}_{,r}+4 )
 +W_{,r}^2+2W W_{,rr} 
 \right. \nonumber \\ 
&& \left. + \frac{2 W \left[W_{,\theta} \left(- 
{F_0}_{,\theta}+3 {F_2}_{,\theta}+3 \cot\theta\right)+ W_{,\theta\theta}\right]+ W_{,\theta}^2}{r^2N} \right\},
\end{eqnarray}

\begin{equation}
\frac{2Ne^{2(F_0+F_1-F_2)}}{\sin^2\theta}{\bf G_{t\phi }}=   W_{,\theta} \left[ {F_0}_{,\theta}-3 \left( 
{F_2}_{,\theta}+\cot\theta\right)\right]- \left\{r N 
\left[W_{,r} \left(-r {F_0}_{,r}+3 r 
{F_2}_{,r}+4\right)+r 
W_{,rr}\right]+W_{,\theta\theta}\right\}  ,
\end{equation}

\begin{eqnarray}
&&  \frac{e^{2F_1}}{N}{\bf G_{rr}}= \frac{({F_0}_{,\theta}+\cot\theta)({F_0}_{,\theta}-{F_1}_{,\theta}+
{F_2}_{,\theta})+{F_2}_{,\theta}({F_2}_{,\theta}-{F_1}_{,\theta})
+{F_0}_{,\theta\theta}+{F_2}_{,\theta\theta}-1+[Nr]'}{r^2 N}+{F_1}_{,r} {F_2}_{,r}\nonumber \\
&& 
%
+\frac{2{F_0}_{,r}+{F_1}_{,r}+{F_2}_{,r}}{r}(1+r)+\frac{e^{2(
{F_2}- {F_0})} \sin ^2\theta (r^2W_{,r}^2N-W_{,\theta}^2 ) }{4 N^2} 
+\frac{N'( 
{F_1}_{,r}+{F_2}_{,r})}{2N},
\end{eqnarray}

\begin{eqnarray}
 && e^{2F_1}r^2{\bf G_{r\theta}}=
\cot\theta (
{F_1}_{,r} -{F_2}_{,r} )-{F_0}_{,r\theta}-{F_2}_{,r\theta}+\frac{{F_0}_{,\theta}+{F_1}_{,\theta}}{r}+\frac{r^2 \sin ^2\theta 
W_{,\theta} W_{,r} e^{2( {F_2}- {F_0})}-N'( {F_0}_{,\theta}-{F_1}_{,\theta})}{2 
N} \nonumber \\ 
&& + {F_0}_{,r} ({F_1}_{,\theta}-{F_0}_{,\theta})+
{F_1}_{,r}({F_0}_{,\theta}+{F_2}_{,\theta} )+{F_2}_{,r}( {F_1}_{,\theta}- {F_2}_{,\theta} ),
\end{eqnarray}

\begin{eqnarray}
&&  e^{2F_1}r^2{\bf G_{\theta\theta}}=r^2 N\left[ {F_0}_{,r}({F_0}_{,r}- 
{F_1}_{,r}+{F_2}_{,r}) + {F_0}_{,rr}+
{F_2}_{,rr}+ {F_2}_{,r}^2-{F_1}_{,r} {F_2}_{,r} \right]
+{F_0}_{,\theta}( {F_1}_{,\theta}+ 
{F_2}_{,\theta})+{F_1}_{,\theta} \
{F_2}_{,\theta}
 \nonumber \\
&& +\cot \theta( {F_0}_{,\theta}+ {F_1}_{,\theta})+\frac{r^2 \sin ^2\theta 
 e^{2( {F_2}- {F_0})}\left(W_{,\theta}^2-Nr^2  W_{,r}^2\right)}{4 N}+r^2 N' \left(\frac{3}{2} {F_0}_{,r}-\frac{1}{2}  {F_1}_{,r}+
{F_2}_{,r}+\frac{1}{r}\right)+\frac{1}{2} r^2 \
N''
 \nonumber \\
&&
+r N(
{F_0}_{,r}-
{F_1}_{,r}+2 {F_2}_{,r}),
\end{eqnarray}

\begin{eqnarray}
&& 4r^2e^{2F_1}{\bf G_{\phi\phi}}= 2 r N'  \left(3 r {F_0}_{,r}+r 
{F_1}_{,r}+2\right)+4 (
{F_0}_{,\theta}^2  +4{F_0}_{,\theta\theta}+ {F_1}_{,\theta\theta}) +2 r^2 N''  \nonumber \\
&& 
+4 r N  \left[r \left({F_0}_{,rr}+{F_1}_{,rr}\right)+r 
{F_0}_{,r}^2+{F_0}_{,r}+{F_1}_{,r}\right] +r^4 \sin ^2\theta e^{2 ({F_2}-F_0)}\left\{ W W_{,r}\left(2  {F_0}_{,r}-6 {F_2}_{,r}-\frac{8}{r}\right)-3  W_{,r}^2\right. \nonumber \\
&&
\left.-2 WW_{,rr}- \frac{\left\{2 W \left[W_{,\theta} \left(- {F_0}_{,\theta}
+3 {F_2}_{,\theta}+3 \cot \theta\right)+ W_{,\theta\theta}\right]+3 W_{,\theta}^2\right\}}{Nr^2}\right\} .
\end{eqnarray}

The components of the Proca energy-momentum tensor~\eqref{procaemt}, for the ansatz~\eqref{procaclouds} and the geometry~\eqref{kerrnc} are also involved because of the four extra functions in the Proca ansatz, and the three new parameters $m,w,\mu$; but they can still be presented in a fairly compact form:

\begin{eqnarray}
&& 2e^{2F_0+2F_1+2F_2}{\bf T_{tt}}=e^{2F_2}\left[ W^2(m{H_1}  
- \sin\theta  
{H_3}_{,r} )^2
-(wH_1+V_{,r})^2\right] \nonumber \\
&&
-\frac{N e^{2 ({F_0}-F_1)}}{r^2} \left\{e^{2 {F_1}}\left[ {\mu^2} r^2 {H_1}^2 e^{2 {F_2}}
+({m} \csc \theta  {H_1}
-{H_3}_{,r})^2\right] +e^{2
{F_2}}\left( {H_1}_{,\theta}- 
{H_2}_{,r}\right)^2\right\}
\nonumber \\
&&
- \frac{e^{2 F_0}}{r^4} \left[{\mu^2} 
r^2( {H_3}^2 e^{2 {F_1}}+ {H_2}^2 e^{2
{F_2}}) + ({H_3}_{,\theta}+ \cot
\theta  {H_3}-mH_2\csc\theta)^2\right]  \nonumber \\
&&
+\frac{ e^{2 (F_1+{F_2})} {\mu^2} r^2\left({H_3}^2  W^2 \sin ^2\theta  - V^2\right) -e^{2 {F_1}}\left(
{m} \csc\theta 
 V +w{H_3}  \right)^2  +e^{2 {F_2}}\left\{W^2[(H_3\sin\theta)_{,\theta}-mH_2]^2-[V_{,\theta}+wH_2]^2\right\}}{r^2N}, \nonumber \\
 && 
%
\end{eqnarray}

\begin{eqnarray}
&&  e^{2F_0+2F_1}{\bf T_{t\phi}}=\frac{[mH_2-(\sin\theta H_3)_{,\theta}][V_{,\theta}+W(H_3\sin\theta)_{,\theta}+H_2(w-mW)]-{\mu^2}r^2 \sin\theta {H_3} e^{2 {F_1}} \left[ 
V+WH_3 \sin\theta\right]}{r^2N}\nonumber \\ 
&&
- \left({m} {H_1}-\sin\theta 
{H_3}_{,r}\right) \left[{H_1} ({m} W-w)-\sin \theta {H_3}_{,r} W-V_{,r}\right],
\end{eqnarray}

\begin{eqnarray}
&& 2r^4e^{2(F_0+F_1+F_2)}{\bf T_{rr}}=
r^2 N e^{2 {F_0}} \left[
e^{2 F_2} {H_1}^2 {\mu^2} r^2
+\left( {m H_1} \csc\theta- {H_3}_{,r}\right)^2+e^{2 ({F_2}-F_1)}( {H_1}_{,\theta}-
{H_2}_{,r})^2\right] \nonumber \\ 
&&
-{\mu^2} r^2e^{2 {F_0}}( {H_3}^2 e^{2 {F_1}}+{H_2}^2e^{2F_2})-e^{2 {F_0}}\csc\theta\left[(\sin\theta{H_3})_{,\theta}-mH_2\right]^2  \nonumber \\
&&
-r^4 e^{2 {F_2}} \left[ {H_1} (w-{m} W) + \sin\theta  {H_3}_{,r} W+  
V_{,r}\right]^2
+\frac{r^2}{N} \left\{e^{2 {F_1}}({m} \csc\theta 
 V+{H_3}w)^2 +{\mu^2} r^2e^{2 ({F_1}+F_2)}[V+ {H_3} W \sin\theta]^2 
 \right. \nonumber \\
&&
\left.
+e^{2 {F_2}}\left[
V_{,\theta}+ (\sin\theta  
{H_3})_{,\theta} W+ 
{H_2} (w-{m} W)\right]^2\right\},
\end{eqnarray}

\begin{eqnarray}
&&  r^2e^{2F_1}{\bf T_{r\theta}}= \frac{ \csc ^2\theta e^{-2 {F_2}}}{r^2} \left({m} {H_1}-\sin \theta {H_3}_{,r}\right) 
 \left[{m} {H_2}-(\sin\theta 
{H_3})_{,\theta}\right] +{\mu^2}  {H_1} 
{H_2}
 \nonumber \\
&&
+\frac{e^{-2 {F_0}}}{N} \left\{[ (w-{m}W) {H_1}+V_{,r}+ W{H_3}_{,r}\sin\theta ] \left[(w-{m}W) {H_2}+V_{,\theta}+W
(\sin\theta{H_3})_{,\theta}\right]
\right\},
\end{eqnarray}

\begin{eqnarray}
&& 2r^4e^{2(F_0+F_1+F_2)}{\bf T_{\theta\theta}}=-r^2 N e^{2 {F_0}} \left[e^{2 F_2} {H_1}^2 
{\mu^2} r^2 +( m\csc\theta H_1-{H_3}_{,r})^2  -e^{2( {F_2}-F_1)}( {H_1}_{,\theta}- 
{H_2}_{,r})^2\right] \nonumber \\
&&
-{\mu^2} r^2e^{2 {F_0}}({H_3}^2 e^{2 {F_1}}-{H_2}^2e^{F_2})
+e^{2 {F_0}} \left(\cot \theta {H_3}-m{H_2} \csc\theta+ {H_3}_{,\theta}\right)^2
 \nonumber \\
&&
+ r^4 e^{2 {F_2}} \left[
 {H_1} (w-{m} W)+ \sin\theta  {H_3}_{,r} W+
V_{,r} \right]^2
+\frac{r^2}{N} \left\{e^{2 {F_1}}({m} \csc\theta 
 V+{H_3}w)^2 +{\mu^2} r^2e^{2 ({F_1}+F_2)}[V+ {H_3} W \sin\theta]^2 
 \right. \nonumber \\
&&
\left.
-e^{2 {F_2}}\left[
V_{,\theta}+ (\sin\theta  
{H_3})_{,\theta} W+ 
{H_2} (w-{m} W)\right]^2\right\},
\end{eqnarray}

  
 \begin{eqnarray}
&& 2e^{2F_0+2F_1+2F_2}{\bf T_{\phi\phi}}=e^{2F_2}\left[ -W^2(m{H_1}  
- \sin\theta  
{H_3}_{,r} )^2
+(wH_1+V_{,r})^2\right] \nonumber \\
&&
-\frac{N e^{2 {F_0}}}{r^2} \left\{\left[ {\mu^2} r^2 {H_1}^2 e^{2 {F_2}}
-({m} \csc \theta  {H_1}
-{H_3}_{,r})^2\right] +e^{2(
{F_2}-F_1)}\left( {H_1}_{,\theta}- 
{H_2}_{,r}\right)^2\right\}
\nonumber \\
&&
+ \frac{e^{2 F_0}}{r^4} \left[{\mu^2} 
r^2( {H_3}^2 e^{2 {F_1}}- {H_2}^2 e^{2
{F_2}}) + ({H_3}_{,\theta}+ \cot
\theta  {H_3}-mH_2\csc\theta)^2\right]  \nonumber \\
&&
+\frac{- e^{2 (F_1+{F_2})} {\mu^2} r^2\left({H_3}^2  W^2 \sin ^2\theta  - V^2\right) -e^{2 {F_1}}\left(
{m} \csc\theta 
 V +w{H_3}  \right)^2  -e^{2 {F_2}}\left\{W^2[(H_3\sin\theta)_{,\theta}-mH_2]^2-[V_{,\theta}+wH_2]^2\right\}}{r^2N} .
\nonumber \\
 && 
\end{eqnarray}  

The four  Proca potentials satisfy the equations\footnote{In deriving this form, we have used
the Proca field equations~\eqref{procafe} together with the Lorentz condition (\ref{lorentz}).}:
\begin{eqnarray}
&&
( r^2 N'-2rN)  W( mH_1
- \sin \theta   {H_3}_{,r} )
-r^2 w N'  {H_1}+\frac{e^{2( {F_1}-F_0)}}{N} V \left[r^2 
(w-{m} W)^2-N e^{2 {F_0}} 
\left({\mu^2} r^2 +{m}^2 \csc 
^2\theta e^{-2F_2}\right)\right]
 \nonumber \\
&&
+[2 (\sin \theta {H_3})_{,\theta} W+V_{,\theta}-2mWH_2)] 
(-{F_0}_{,\theta}+{F_2}_{,\theta}+\cot\theta) -{H_2} 
\left[
2 w {F_0}_{,\theta}+{m} W_{,\theta}\right] +(\sin \theta  {H_3})_{,\theta} W_{,\theta}
+ V_{,\theta\theta}
 \nonumber \\
&&
+r N  \left\{
 [2 W(mH_1- \sin \theta   {H_3}_{,r})-V_{,r}] \left(r {F_0}_{,r}-r 
{F_2}_{,r}\right)
-rH_1 \left(2 w {F_0}_{,r}+{m} 
W_{,r}\right)
+r \sin \theta {H_3}_{,r} W_{,r}+2 
V_{,r}+r V_{,rr}\right\}
 \nonumber \\
&&
+r^2 \sin ^2\theta e^{2 ({F_2}-F_0)}
W\left\{
 W_{,\theta} \frac{\left[{H_2} (w-{m} W)+(\sin \theta
{H_3})_{,\theta} W+V_{,\theta}\right]}{N}+r^2 W_{,r}[
H_1(w-{m}  W) 
+ \sin\theta  {H_3}_{,r} W
+ V_{,r} ]
\right\}
=0, \nonumber \\
&&
\end{eqnarray}

\begin{eqnarray}
&&  r^3 [{H_1} N]'' + r^2 [{H_1} N]'  \left(r {F_0}_{,r}-2 r {F_1}_{,r}+r 
{F_2}_{,r}+2\right)
-r N  {H_1} \left[2 r {F_1}_{,r} \left(r 
{F_0}_{,r}+r {F_2}_{,r}+2\right)-r^2( 
{F_0}_{,rr}+ {F_2}_{,rr})+2\right]\nonumber \\
&&
+[r {H_1}_{,\theta}-2 {H_2}\left(r {F_1}_{,r}+1\right)] ( {F_0}_{,\theta}+ {F_2}_{,\theta}+\cot\theta )
+rH_2( {F_0}_{,r\theta}+
 {F_2}_{,r\theta})
 -r {H_1}e^{2 {F_1}} \left({\mu^2} r^2 +e^{-2F_2}{m}^2 \csc ^2\theta\right)
   \nonumber \\
&&
-2( r {F_1}_{,r}+1) {H_2}_{,\theta}
+2 r {F_1}_{,\theta} ({H_2}_{,r} -{H_1}_{,\theta})
+r  {H_1}_{,\theta\theta} 
+2 {m}  \csc\theta e^{2 
({F_1}-F_2)}{H_3} (r {F_2}_{,r}+1) 
 \nonumber \\
&&
+\frac{r^3  e^{2 
({F_1}-F_0)}}{N} \left\{
\left(2{F_0}_{,r}+\frac{N'}{N}\right)(w-mW)(V+\sin \theta {H_3} 
W )
+W_{,r}[ \sin \theta {H_3}  
(W {m}-w)+m(V+\sin\theta H_3W)] 
\right\} =0, \nonumber \\
&&
\end{eqnarray}

  
\begin{eqnarray}   
&& 
r^2 [N {H_2}_{,r}]'
-2 r^2 {F_1}_{,\theta} [N{H_1}]'
+{H_2}_{,\theta\theta}
+ {H_2}_{,\theta} \left({F_0}_{,\theta}-2 {F_1}_{,\theta}+{F_2}_{,\theta}+\cot\theta\right)
+2 {m} \csc \theta e^{2 ({F_1}-F_2)} {H_3} \left({F_2}_{,\theta}
+\cot \theta\right) \nonumber \\
&&
+r^2 N {H_2}_{,r} \left[{F_0}_{,r} 
-2 {F_1}_{,r} 
+{F_2}_{,r} \right]
-rN{H_1} \left[2 {F_1}_{,\theta} \left(r {F_0}_{,r}+r {F_2}_{,r}+2\right) 
-r \left({F_0}_{,r\theta}+{F_2}_{,r\theta}\right)\right]
+2rN \left(r {F_1}_{,r}+1\right) {H_1}_{,\theta}
\nonumber \\
&&
- H_2 \left[2 {F_1}_{,\theta} 
\left({F_0}_{,\theta}+{F_2}_{,\theta}+\cot \theta\right)
-{F_0}_{,\theta\theta} +{\mu^2} r^2 e^{2 
{F_1}}+{m}^2 \csc ^2\theta e^{2( 
{F_1}-F_2)}- {F_2}_{,\theta\theta}+\csc ^2\theta \right] \nonumber \\
&&
+\frac{r^2 e^{2 ({F_1}-F_0)} }{N}
\left\{ {H_2} (w-{m} W)^2 
+2{F_0}_{,\theta}(w-mW)(V+\sin\theta H_3W)+W_{,\theta}[\sin \theta {H_3}(Wm-w)+m(V+\sin \theta {H_3}W)]\right\} =0,  \nonumber \\
&&
\end{eqnarray}

\begin{eqnarray}
&&
r^2 [N  {H_3}_{,r}]'
+{H_3}_{,\theta\theta}+ {H_3} 
 \left[
  \cot \theta ({F_0}_{,\theta} -{F_2}_{,\theta})
 - {\mu^2} r^2 e^{2 {F_1}}
 - {m}^2 \csc ^2\theta e^{2( 
{F_1}-F_2)}
-\csc ^2\theta 
\right] 
+{H_3}_{,\theta} \left({F_0}_{,\theta}-{F_2}_{,\theta}+\cot \theta\right)
  \nonumber \\
&&
+\sin\theta r^4e^{2 ({F_2}-F_0)}W_{,r}[
   {H_1}(m W 
- w)   
-( \sin \theta  {H_3}_{,r} W 
+ V_{,r}) ]
+2 {m} \csc \theta [
{H_2} 
\left({F_2}_{,\theta}+\cot \theta\right) 
+ rN 
\left(r {F_2}_{,r}+1\right) {H_1}]
\nonumber \\
&&
+ r^2N{H_3}_{,r} \left({F_0}_{,r} 
-{F_2}_{,r}\right)
+\frac{ r^2e^{-2F_0} }{N} 
\left\{
{H_3}e^{2 {F_1}} \left({m}  W-w \right)^2 
- e^{2 {F_2}}\sin\theta W_{,\theta} [
 {H_2} (w-{m} W)+(\sin \theta
{H_3})_{,\theta} W+V_{,\theta}]\right\}=0 .
\nonumber \\
&&
\end{eqnarray}

For completeness, we also exhibit the gauge condition~\eqref{lorentz}, which reads:

\begin{eqnarray}
&&
e^{-2 {F_1}} \left\{r^2 N{H_1}( {F_0}_{,r}+ {F_2}_{,r})
+({F_0}_{,\theta} + {F_2}_{,\theta} ){H_2}
+[r^2 H  {H_1}]'
+ {H_2}_{,\theta}
+\cot \theta {H_2}
\right\}
 \nonumber \\
&&
+\frac{r^2e^{-2 {F_0}}}{N}
 (\sin \theta  {H_3} W+  V)(mW-w)
-{m} \csc \theta e^{-2 {F_2}} {H_3}=0 \ .
\end{eqnarray}

\section{Angular momentum -- Noether charge relation and 
a simplified Smarr formula}
\label{appendixc}
As discussed in the main text, the relation $J = m Q$ [equation~\eqref{amnc}] holds for Proca stars, in analogy with the case of scalar boson stars. In contrast with the latter, however, the angular momentum \textit{density} and Noether charge \textit{density} are $not$ proportional; the proportionality only holds at the level of the integrated quantities, but with the further subtlety of a possible boundary term if a horizon is present. This is an interesting distinction between the two types of stars (and hairy BHs) which could not be anticipated.

\begin{figure}[h!]
  \begin{center}
    \includegraphics[width=8.1cm]{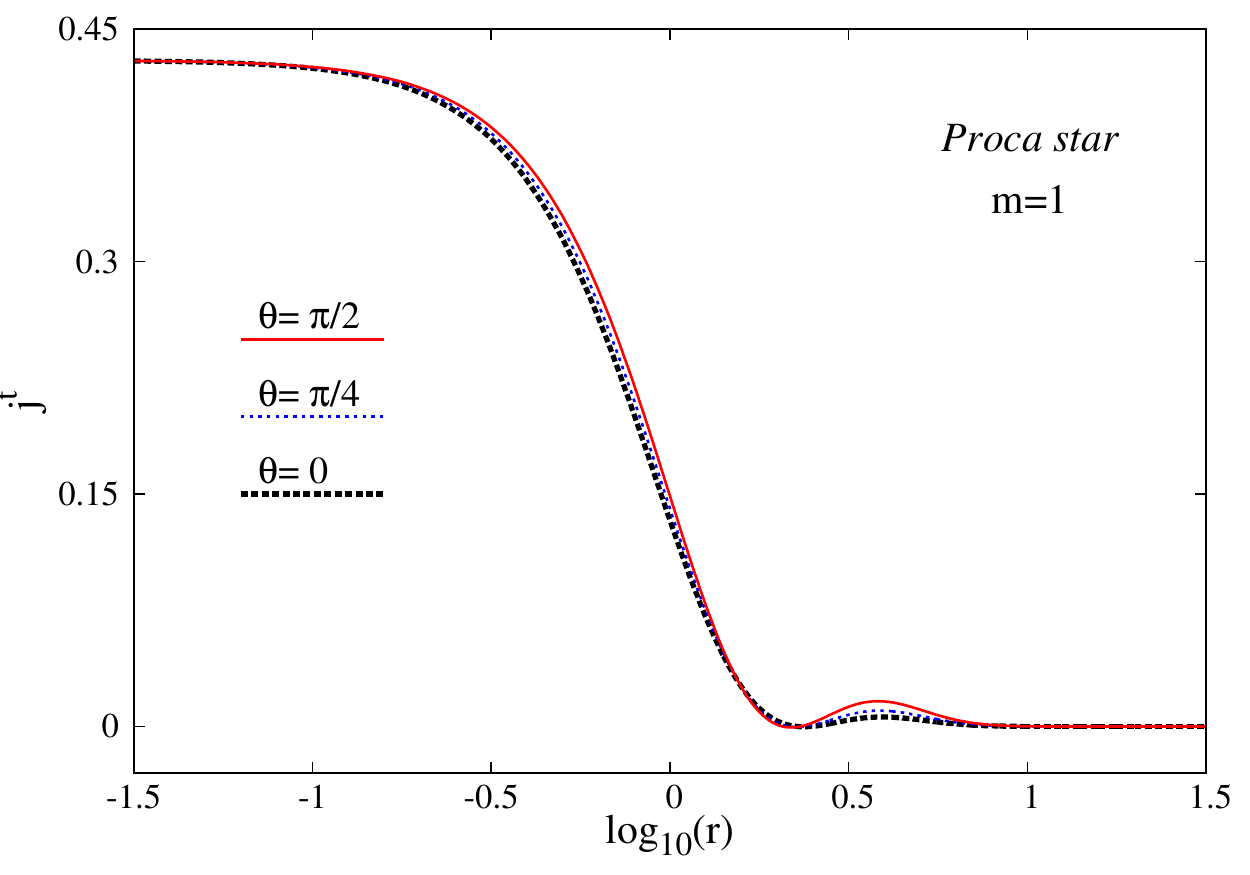}
      \includegraphics[width=8.1cm]{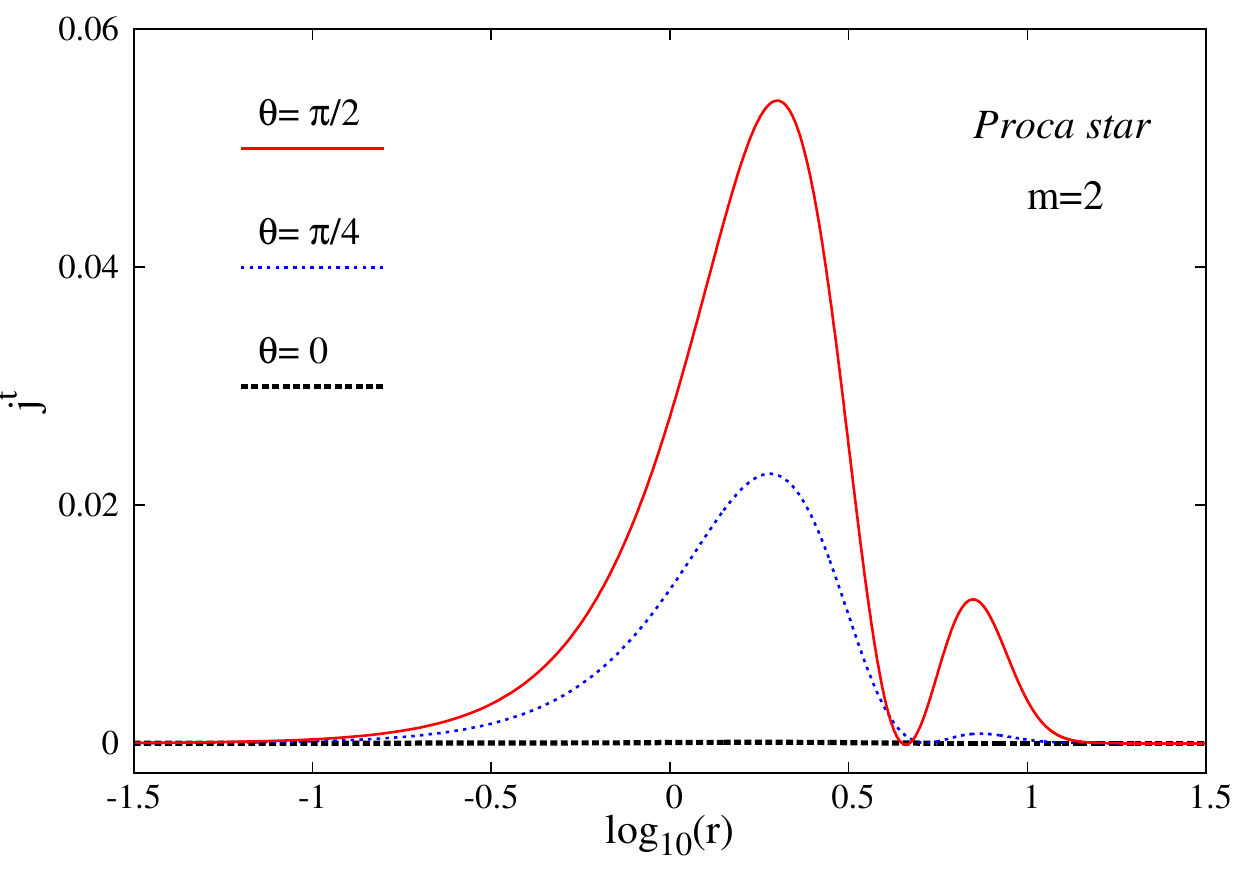}
  \end{center}
  \caption{Radial variation of the Noether charge density, $cf.$~\eqref{j}, for different constant $\theta$ sections of the Proca stars exhibited in Fig.~\ref{PS1} (left panel) and Fig.~\ref{PS2} (right panel). One can clearly observe the differences with the angular momentum density.}
  \label{fignc}
\end{figure}

To prove the relation~\eqref{amnc}, one starts with the  expression of the angular momentum density:
\begin{eqnarray}
\label{s1}
T_\varphi^t=\frac{1}{2}
\left [
 {\mathcal{F}}_{\alpha \varphi} \bar{\mathcal{F}}^{\alpha t}+\bar{\mathcal{F}}_{\alpha \varphi}  {\mathcal{F}}^{\alpha t}
 +\mu^2 (
\mathcal{A}_{\varphi} \bar{\mathcal{A}}^{t}
 +\bar{\mathcal{A}}_{\varphi}\mathcal{A}^{t}
)
\right ]  \ ,
\end{eqnarray}
where one introduces the expressions  
\begin{eqnarray}
\label{s2}
 {\mathcal{F}}_{\alpha \varphi} =\partial_\alpha {\mathcal{A}}_{\varphi}-i m {\mathcal{A}}_{\alpha}\ , \qquad 
\bar{ {\mathcal{F}}}_{\alpha \varphi} =\partial_\alpha \bar{{\mathcal{A}}}_{\varphi}+i m \bar{{\mathcal{A}}}_{\alpha} \ .
\end{eqnarray}
Thus
\begin{eqnarray}
\label{s3}
T_\varphi^t  
=\frac{1}{2}
\left\{
\frac{1}{\sqrt{-g}}\partial_\alpha \left[ ({\mathcal{A}}_\varphi \bar{ {\mathcal{F}}}^{\alpha t}+\bar{\mathcal{A}}_\varphi { {\mathcal{F}}}^{\alpha t}  )\sqrt{-g} \right]
-i m 
(
\mathcal{A}_{\alpha} \bar{ {\mathcal{F}}}^{\alpha t}
- 
 \bar{\mathcal{A}}_{\alpha}{ {\mathcal{F}}}^{\alpha t}
)
\right.
\nonumber 
\\
\left. - \frac{1}{\sqrt{-g}}\left[{\mathcal{A}}_\varphi\partial_\alpha( \bar{ {\mathcal{F}}}^{\alpha t}\sqrt{-g})
+\bar{\mathcal{A}}_\varphi \partial_\alpha({ {\mathcal{F}}}^{\alpha t} \sqrt{-g}) \right] 
+\mu^2 (
\mathcal{A}_{\varphi} \bar{\mathcal{A}}^{t}
 +\bar{\mathcal{A}}_{\varphi}\mathcal{A}^{t}
)
\right\} \ .
\end{eqnarray}
The above expression can be simplified by using the Proca equations (\ref{procafe}).
Then the second line is identically zero and one arrives at the expression
[with $j^t$ defined  in (\ref{j})]
\begin{eqnarray}
\label{s4}
T_\varphi^t  =m j^t 
+ \nabla_\alpha P^\alpha \ ,
\end{eqnarray}
with
\begin{eqnarray}
\label{s5}
  P^\alpha= {\mathcal{A}}_\varphi \bar{ {\mathcal{F}}}^{\alpha t}+\bar{\mathcal{A}}_\varphi { {\mathcal{F}}}^{\alpha t}   \ .
\end{eqnarray}
Thus, the angular momentum density and Noether charge density (multiplied by the azimuthal harmonic index $m$) differ by a total divergence. We have checked that $\nabla_\alpha P^\alpha$ is locally nonzero, as illustrated in Fig.~\ref{fignc}. 

For BHs, on the other hand, there is also a horizon contribution which, in general is non-zero:
\begin{equation}
J^{(\mathcal{P})}=mQ+ \oint_\mathcal{H}  P^r dS_r\ .
\label{JQBHs}
\end{equation}  

This result  contrasts with that for the scalar boson star case, wherein $T_\varphi^t =m j^t $. The total angular momentum and Noether charge, however, are equal in both cases, for stars. This follows from integrating (\ref{s4}) over a spacelike surface and using the Gauss's theorem:
\begin{eqnarray}
\label{s6}
\int T_\varphi^t \sqrt{-g} d^3 x =m \int j^t \sqrt{-g} d^3 x
+ \oint_\infty  P^r dS_r\ .
\end{eqnarray}
Since the Proca field decays exponentially, the contribution from the $P^r$ term 
is zero and one arrives at (\ref{amnc}).

By using a similar approach, one can easily prove the following identity, where use again the Proca equations (\ref{procafe}) together with
the expressions 
$ {\mathcal{F}}_{\alpha t} =\partial_\alpha {\mathcal{A}}_{t}+i w {\mathcal{A}}_{t}$,
$ \bar{\mathcal{F}}_{\alpha t} =\partial_\alpha \bar{\mathcal{A}}_{t}-i w \bar {\mathcal{A}}_{t}$:
 \begin{equation}
\label{sup1}
T^{\alpha}_\alpha-2T_t^t=
2w j^t - \mu^2 {\mathcal{A}}_\alpha \bar{\mathcal{A}}^\alpha+
\nabla_\alpha U^\alpha,
\end{equation} 
 with
\begin{eqnarray}
\label{sup2}
 U^\alpha=\frac{1}{2} 
\left(
{\mathcal{A}}_\beta \bar{ {\mathcal{F}}}^{\alpha \beta}+\bar{\mathcal{A}}_\beta { {\mathcal{F}}}^{\alpha \beta}
\right)
-\left({\mathcal{A}}_t \bar{ {\mathcal{F}}}^{\alpha t}+\bar{\mathcal{A}}_t { {\mathcal{F}}}^{\alpha t}
\right)  ~.
\end{eqnarray}
For Proca stars,
the integration of the relation (\ref{sup1}) 
over a spacelike surface 
implies the Smarr formula (\ref{smarr-new2}). Here we use again the fact that the Proca field decays exponentially, while the contribution from  $U^r$ at $r=0$ vanishes.

For BHs, however, $U^r$ gives a nontrivial horizon contribution. Then, a combination of (\ref{s4}) and (\ref{sup1}) leads to the simple relation
\begin{eqnarray}
\label{sup3}
\nonumber
&&T^{\alpha}_\alpha-2T_t^t-2\Omega_H T_\varphi^t =	 2(w-\Omega_H m) j^t - \mu^2 {\mathcal{A}}_\alpha \bar{\mathcal{A}}^\alpha \ \ \ \ \ \ \ \ \ \ \ 
\\
&&+
\frac{1}{\sqrt{-g}}
\partial_\alpha 
\bigg[\left(\frac{1}{2} 
\left(
{\mathcal{A}}_\beta \bar{ {\mathcal{F}}}^{\alpha \beta}+\bar{\mathcal{A}}_\beta { {\mathcal{F}}}^{\alpha \beta}
\right)
-\left(
({\mathcal{A}}_t +\Omega_H {\mathcal{A}}_\varphi)\bar{ {\mathcal{F}}}^{\alpha t}
                +
(\bar{\mathcal{A}}_t +\Omega_H \bar{\mathcal{A}}_\varphi){ {\mathcal{F}}}^{\alpha t}
\right)
\right)  
\sqrt{-g} 
\bigg] \ .
\end{eqnarray}
The integration of this identity leads to the simplified 
Smarr formula (\ref{smarr-new1}). Here one uses also the synchronization 
condition (\ref{synchronization}), 
together with  the boundary conditions (\ref{bccloudshorizon}).

\section{Numerical data made available}
\label{appendixd}
The following tables detail the information about the four solutions singled out in Fig.~\ref{figdomain}, for each case (Proca and scalar), plus a vacuum Kerr solution that possesses the same ADM mass and angular momentum as configuration III. The numerical data, together with an explanation of its format, is given in  \cite{datakbhph} for the Proca case and in \cite{datakbhsh} for the scalar case.

\bigskip

\begin{center}
\begin{tabular}{|c||c|c|c|c|c|c|c|c|}
\hline
Left panel Fig.~\ref{figdomain} & $w/\mu$  & $r_H/\mu$ & $\mu M_{ADM}$ & $\mu^2J_{ADM}$ & $\mu M_{H}$ & $\mu^2J_{H}$ & $\mu M^{(\mathcal{P})}$ & $\mu^2J^{(\mathcal{P})}$   \\ \hline\hline
I - Proca star  & 0.9 & 0 & 1.456 & 1.45 & 0 & 0 & 1.456 & 1.45 
\\ \hline
II - Vacuum Kerr  & 1.0432 & 0.1945 & 0.365 & 0.128 & 0.365 & 0.128 & 0 & 0
\\ \hline
III - KBHPH & 0.9775 & 0.2475 & 0.365 & 0.128 & 0.354 & 0.117 & 0.011 & 0.011
\\ \hline
IV - KBHPH & 0.863& 0.09 & 0.915 & 0.732 & 0.164 & 0.070 & 0.751 & 0.662
\\ \hline
V - KBHPH & 0.79& 0.06 & 1.173 & 1.079 & 0.035 & 0.006 & 1.138 & 1.073
\\ \hline 
\end{tabular}\\ \bigskip
Table I - Configurations in data publicly available for the Proca case~\cite{datakbhph}.
\end{center}

\bigskip

\begin{center}
\begin{tabular}{|c||c|c|c|c|c|c|c|c|}
\hline
Left panel Fig.~\ref{figdomain} & $w/\mu$  & $r_H/\mu$ & $\mu M_{ADM}$ & $\mu^2J_{ADM}$ & $\mu M_{H}$ & $\mu^2J_{H}$ & $\mu M^{(\Psi)}$ & $\mu^2J^{(\Psi)}$   \\ \hline\hline
I - Scalar boson star  & 0.85 & 0 & 1.25 & 1.30 & 0 & 0 & 1.25 & 1.30 
\\ \hline
II - Vacuum Kerr  & 1.1112 & 0.0663 & 0.415 & 0.172 & 0.415 & 0.172 & 0 & 0
\\ \hline
III - KBHSH & 0.975 & 0.2 & 0.415 & 0.172 & 0.393 & 0.150 & 0.022 & 0.022
\\ \hline
IV - KBHSH & 0.82& 0.1 & 0.933 & 0.739 & 0.234 & 0.114 & 0.699 & 0.625
\\ \hline
V - KBHSH & 0.68& 0.04 & 0.975 & 0.850 & 0.018 & 0.002 & 0.957 & 0.848
\\ \hline 
\end{tabular}\\ \bigskip
Table II - Configurations in data publicly available for the scalar case~\cite{datakbhsh}.
\end{center}

\begin{small}

 \end{small}

\end{document}